\documentclass[aps,preprint,prd,nofootinbib]{revtex4}
\usepackage{graphicx}
\usepackage{psfrag}

\begin{document}

\title{Spectrum for Heavy Quankonia and Mixture of the Relevant Wave Functions
within the Framework of Bethe-Salpeter Equation} \vspace{8mm}
\author{Chao-Hsi Chang $^{1,2,3,4}$\footnote{email:zhangzx@itp.ac.cn}, Guo-Li Wang$^1$
\footnote{email:gl\_wang@hit.edu.cn}}
\address{
$^1$ Department of Physics, Harbin
Institute of Technology, Harbin, 150001, China\\
$^2$ CCAST (World Laboratory), P.O.Box 8730, Beijing 100190, P.R.
China.\\
$^3$ Institute of Theoretical Physics, Chinese Academy of Sciences,
P.O.Box 2735, Beijing 100190, P.R. China.\\
$^4$ Department of Physics, Chongqing University, Chongqing 400044,
P.R. China }
\baselineskip=20pt

\begin{abstract}
Considering the fact that some excited states of the heavy quarkonia
(charmonium and bottomonium) still missing in experimental
observations and potential applications of the relevant wave
functions of the bound states, we re-analyze the spectrum and the
relevant wave functions of the heavy quarkonia within the framework
of Bethe-Salpeter (B.S.) equation with a proper QCD-inspired kernel.
Such a kernel for the heavy quarkonia, relating to potential of
non-relativistic quark model, is instantaneous, so we call the
corresponding B.S. equation as BS-In equation throughout the paper.
Particularly, a new way to solve the B.S. equation, which is
different from the traditional ones, is proposed here, and with it
not only the known spectrum for the heavy quarkonia is re-generated,
but also an important issue is brought in, i.e., the obtained
solutions of the equation `automatically' include the `fine',
`hyperfine'~splittings and the wave function mixture, such as $S-D$
wave mixing in $J^{PC}=1^{--}$ states, $P-F$ wave mixing in
$J^{PC}=2^{++}$ states for charmonium and bottomonium etc. It is
pointed out that the best place to test the wave mixture probably is
at $Z$-factory ($e^+e^-$ collider running at $Z$-boson pole with
extremely high luminosity).
\end{abstract}

\maketitle

\section{Introduction}

Spectroscopy, including the spectrum and the corresponding wave
functions, is a very interesting topic for heavy quarkonia in
particle physics. The spectrum and the corresponding wave functions
for the binding systems can be tested experimentally and via study
of the spectroscopy one may have insight of the heavy quarkonia and
understand QCD , which is the nature of the binding, further as
well. In the literature, there are various approaches to the
spectroscopy of the heavy quarkonia: charmonium and bottomonium
\cite{Eichten1,Eichten2,Godfrey,ypk,ktchao,caix,rob}, and to solve
the Bethe-Salpeter (B.S.) equation is one of them
\cite{ypk,ktchao,caix,rob}. Since recently we have realized a new
method to solve the B.S. equation for the heavy quarkonia, so in
this paper, we would like to try the method i.e. to apply this
method to re-analyzing the spectroscopy of the heavy quarkoia:
charmonium and bottomonium under B.S. equation approach.

First of all, how to determine the B.S. kernel is crucial for B.S.
equation approach to a bound state problem. It is known that if one
adopts the QCD-inspired Bethe-Salpeter (B.S.) equation \cite{BS}
approach to the problems of hadronic bound states, then the relevant
B.S. kernel for a double heavy quark-antiquark system, such as
charmonium and bottomonium, is instantaneous approximately i.e. the
B.S. equation is essentially an instantaneous one (a BS-In
equation). It is also known that BS-In equation can further
precisely relate to the Sch\"{o}dinger equation in potential model
(PM) by means of the Salpeter approximate method \cite{salp}.
Therefore, one may use the relation to the potential model (PM) and
help oneself to determine the kernel of BS-In equation precisely.
Whereas starting with the BS-In equation whose kernel is fixed in
terms of QCD consideration and the relation to PM, one can extend
some relativistic nature of the problem more than what PM can
consider, and stand on more solid theoretical ground for the B.S.
equation approach, hence, we start the study of spectroscopy for
heavy quarkia with such a BS-In equation. Moreover, we apply the new
realized method to solving the BS-In equation. People later on will
see an important issue from the new method is that besides the `fine
and hyperfine' splitting being involved, the wave mixtures in the
wave functions, such as $S-D$ wave mixing in $J^{PC}=1^{--}$ ($J$:
total angle momentum; $P$: parity; $C$: charge parity) states and
$P-F$ wave mixing in $J^{PC}=2^{++}$ states etc, are determined
precisely, although the mixtures, in fact, are rooted in the kernel
under the present framework of BS-In equation.

The new proposed method can be outlined as that, firstly we analyze
the bound states according to their total angular momentum $J$,
parity $P$ and charge conjugation $C$, such as the states
$0^{-+}$($^1S_0$), $1^{--}$($^3S_1$ or $^3D_1$), $0^{++}$($^3P_0$),
$1^{++}$($^3P_1$), $2^{++}$($^3P_2$ or $^3F_2$), and
$1^{+-}$($^1P_1$) etc; secondly we write down the most general
formulation for the B.S. wave functions respectively, and then input
the formulated wave functions into the BS-In equation and turn the
equation into a set of proper coupled equations for the components
which appear in the formulation; finally we solve the coupled
equation numerically, and obtain the mass spectra and wave functions
for ($c \bar c$) and ($b \bar b$) binding systems. For convenience,
we call the coupled equations as BS-CoEqs later on.

This paper is organized as following, additional to the Introduction
section I, in section II we introduce the relativistic
Bethe-Salpeter equation and BS-In equation. In section III we start
with the generalized formulation for relativistic wave functions
with definite quantum numbers to derive the relevant BS-CoEqs for
low total angle momentum states individually. Finally, we show the
numerical results although we do not present the detail to solve the
equation numerically, and we also explain and briefly discuss the
obtained solutions of the BS-CoEqs in section IV.

\section{Instantaneous Bethe-Salpeter Equation}

Firstly let us outline the reduction of the B.S. equation which is
similar to the way of Salpeter \cite{salp} if B.S. kernel is
instantaneous, and introduce necessary notations. The readers, who
are interested in the details, can also find them in Ref.
\cite{chenjk,cskimwang,changwang}.

The Bethe-Salpeter (B.S.) equation for mesons is read as:
\begin{equation}
(\not\!{p_{1}}-m_{1})\chi(q)(\not\!{p_{2}}+m_{2})=
i\int\frac{d^{4}k}{(2\pi)^{4}}V(P,k,q)\chi(k)\;, \label{eq1}
\end{equation}
where $\chi(q)$ is the B.S. wave function, $V(P,k,q)$ is the
interaction kernel between the quark and antiquark, and $p_{1},
p_{2}$ are the momenta of the quark 1 and anti-quark 2. Quark mass
is $m_1$, antiquark mass is $m_2$, and here we consider heavy
quarkonia: charmonium and bottomonium so we have $m_1=m_2$. The
total momentum $P$ and the relative momentum $q$ are defined as:
$$p_{1}={\alpha}_{1}P+q, \;\; {\alpha}_{1}=\frac{m_{1}}{m_{1}+m_{2}} ~,$$
$$p_{2}={\alpha}_{2}P-q, \;\; {\alpha}_{2}=\frac{m_{2}}{m_{1}+m_{2}} ~,$$
and $\alpha_1=\alpha_2=\frac{1}{2}$ for charmonium and bottomonium.

We divide the relative momentum $q$ into two parts,
$q_{\parallel}$ and $q_{\perp}$,
$$q^{\mu}=q^{\mu}_{\parallel}+q^{\mu}_{\perp}\;,$$
$$q^{\mu}_{\parallel}\equiv (P\cdot q/M^{2})P^{\mu}\;,\;\;\;
q^{\mu}_{\perp}\equiv q^{\mu}-q^{\mu}_{\parallel}\;.$$
Correspondingly, we may have two Lorentz invariant variables:
\begin{center}
$q_{_P}=\frac{(P\cdot q)}{M}\;, \;\;\;\;\;
q_{_T}=\sqrt{q^{2}_{_P}-q^{2}}=\sqrt{-q^{2}_{\perp}}\;.$
\end{center}
When $\stackrel{\rightarrow}{P}=0$, they turn to the usual component
$q_{0}$ and $|\vec q|$ respectively.

If the kernel $V(P,k,q)$ takes the simple form:
$$V(P,k,q) \Rightarrow V(k_{\perp},q_{\perp})\;$$
namely the B.S. equation is `instantaneous', for convenience, we
would like to introduce the notations $\varphi_{p}(q^{\mu}_{\perp})$
and $\eta(q^{\mu}_{\perp})$ so the `instantaneous (three
dimensional) objects' will accordingly read as follows:
\begin{eqnarray} \label{eq05}
\varphi_{_P}(q^{\mu}_{\perp})\equiv i\int
\frac{dq_{_P}}{2\pi}\chi(q^{\mu}_{\parallel},q^{\mu}_{\perp})\;,
\end{eqnarray}
\begin{eqnarray}
\eta(q^{\mu}_{\perp})\equiv\int\frac{dk_{\perp}}{(2\pi)^{3}}
V(k_{\perp},q_{\perp})\varphi_{_P}(k^{\mu}_{\perp})\;. \label{eq5}
\end{eqnarray}
The B.S. equation now is rewritten as:
\begin{equation}
\chi(q_{\parallel},q_{\perp})=S_{1}(p_{1})\eta(q_{\perp})S_{2}(p_{2})\;.
\label{eq6}
\end{equation}
Generally the propagators of the two constituents can be decomposed
as:
\begin{equation}
S_{i}(p_{i})=\frac{\Lambda^{+}_{iP}(q_{\perp})}{J(i)q_{_P}
+\alpha_{i}M-\omega_{i}+i\epsilon}+
\frac{\Lambda^{-}_{iP}(q_{\perp})}{J(i)q_{_P}+\alpha_{i}M+\omega_{i}-i\epsilon}\;,
\label{eq7}
\end{equation}
with
\begin{equation}
\omega_{i}=\sqrt{m_{i}^{2}+q^{2}_{_T}}\;,\;\;\;
\Lambda^{\pm}_{iP}(q_{\perp})= \frac{1}{2\omega_{i}}\left[
\frac{\not\!{P}}{M}\omega_{i}\pm
J(i)(m_{i}+{\not\!q}_{\perp})\right]\;, \label{eq8}
\end{equation}
where $i=1, 2$ for quark and anti-quark respectively, and
$J(i)=(-1)^{i+1}$, and $\Lambda^{\pm}_{iP}(q_{\perp})$ satisfy the
relations:
\begin{equation}
\Lambda^{+}_{iP}(q_{\perp})+\Lambda^{-}_{iP}(q_{\perp})=\frac{\not\!{P}}{M}~,\;\;
\Lambda^{\pm}_{iP}(q_{\perp})\frac{\not\!{P}}{M}
\Lambda^{\pm}_{iP}(q_{\perp})=\Lambda^{\pm}_{iP}(q_{\perp})~,\;\;
\Lambda^{\pm}_{iP}(q_{\perp})\frac{\not\!{P}}{M}
\Lambda^{\mp}_{iP}(q_{\perp})=0~. \label{eq9}
\end{equation}
Hence sometimes $\Lambda^{\pm}_{iP}(q_{\perp})$ are called as
`project operators', although they need to be sandwiched with the
operator $\frac{\not\!{P}}{M}$ when `projecting' as Eq(\ref{eq9}).

Introducing the notations $\varphi^{\pm\pm}_{_P}(q_{\perp})$ to note
the projected wave functions as:
\begin{equation}
\varphi^{\pm\pm}_{_P}(q_{\perp})\equiv \Lambda^{\pm}_{1P}(q_{\perp})
\frac{\not\!{P}}{M}\varphi_{_P}(q^\mu_{\perp}) \frac{\not\!{P}}{M}
\Lambda^{{\pm}}_{2P}(q_{\perp})\;, \label{eq10}
\end{equation}
and we indeed have
$$
\varphi_{_P}(q^\mu_{\perp})=\varphi^{++}_{_P}(q^\mu_{\perp})+
\varphi^{+-}_{_P}(q^\mu_{\perp})+\varphi^{-+}_{_P}(q^\mu_{\perp})
+\varphi^{--}_{_P}(q^\mu_{\perp})
$$
With contour integration over $q_{p}$ on both sides of
Eq.(\ref{eq6}), we obtain:
$$\varphi_{_P}(q_{\perp})=\frac{
\Lambda^{+}_{1P}(q_{\perp})\eta(q_{\perp})\Lambda^{+}_{2P}(q_{\perp})}
{(M-2\omega_{1})}- \frac{
\Lambda^{-}_{1P}(q_{\perp})\eta(q_{\perp})\Lambda^{-}_{2P}(q_{\perp})}
{(M+2\omega_{1})}\;,
$$
and the equation becomes four independent equations:
\begin{eqnarray}
& (M-2\omega_{1})\varphi^{++}_{_P}(q_{\perp})=
\Lambda^{+}_{1P}(q_{\perp})\eta(q_{\perp})
\Lambda^{+}_{2P}(q_{\perp})\;,\nonumber\\
&(M+2\omega_{1})\varphi^{--}_{_P}(q_{\perp})=-
\Lambda^{-}_{1P}(q_{\perp})\eta(q_{\perp})\Lambda^{-}_{2P}(q_{\perp})\;,\nonumber\\
&\varphi^{+-}_{_P}(q_{\perp})=\varphi^{-+}_{_P}(q_{\perp})=0\;.
\label{eq11}
\end{eqnarray}
where we have $\omega_{1}=\omega_{2}$ for the equal mass system. In
fact the four equations is of an `eigenvalue problem' about the
eigenvalue $M$. Note that in the Ref.\cite{salp} the way for solving
the BS-In equation is not exactly equivalent to the four equations
Eq.(\ref{eq11}). Details about examining the equivalence may be
found in Ref.\cite{chenjk}. Alternately here we exactly start with
the four equations to solve the BS-In equation.

The normalization condition for B.S. wave function is:
\begin{equation}
\int\frac{q_{_T}^2dq_{_T}}{2{\pi}^2}Tr\left[\overline\varphi^{++}
\frac{{/}\!\!\!
{P}}{M}\varphi^{++}\frac{{/}\!\!\!{P}}{M}-\overline\varphi^{--}
\frac{{/}\!\!\! {P}}{M}\varphi^{--}\frac{{/}\!\!\!
{P}}{M}\right]=2P_{0}\;. \label{eq12}
\end{equation}

Now let us return to the problem for the heavy quarkonia
($c\bar{c}$) and ($b\bar{b}$). To fix the kernel for the heavy quark
and heavy anti-quark, on one hand, we should let the kernel being
QCD-inspired and on the other hand, we should relate the kernel to
the Cornell potential accordingly. Thus the kernel in space-time
looks like as a linear scalar interaction (the confinement one in
QCD nonperturbative nature) plus a vector interaction (single gluon
exchange in Coulomb gauge):
\begin{equation}
V(r)=V_s(r)+\gamma_{_0}\otimes\gamma^0 V_v(r)= \lambda
r+V_0-\gamma_{_0}\otimes\gamma^0 \frac{4}{3}\frac{\alpha_s}{r}\;,
\label{eq13}
\end{equation}
where $\lambda$ is the string constant, $\alpha_s(r)$ is the running
coupling constant of QCD. Usually, in order to fit the data of heavy
quarkonia, a constant $V_0$ is often added to the scalar confining
potential and takes different values for the bound states with
different quantum numbers respectively\footnote{One will see later
on in this paper that the value of $V_0$ is determined by fitting
the data for the ground states with the corresponding quantum
numbers.}.

To avoid the infrared divergence in the Coulomb-like one and to
correspond the fact that the confined linear interaction should be
also suppressed at large distance phenomenologically, so it will be
better to re-formulate the kernel as follows:
\begin{eqnarray}
&V_s(r)=\frac{\lambda}{\alpha}(1-e^{-\alpha' r})+V_0~,\nonumber\\
&V_v(r)=-\frac{4}{3}\frac{\alpha_s}{r}e^{-\alpha r}\;. \label{eq14}
\end{eqnarray}
To decrease the parameters which are needed to fix by fitting data,
we assume $\alpha'=\alpha$ approximately\footnote{In fact, at final
step (numerical solving BS-CoEq) we find that the results are not
very sensitive to the assumption when $\alpha$ and $\alpha'$ vary in
reasonable region.}. It is easy to show that when $\alpha r\ll 1$,
the potential approximately becomes linear. Now the B.S. kernel in
momentum space and in the rest frame of the bound state is read as:
\begin{eqnarray}\label{eq-poten}
&V(\stackrel{\rightarrow}{q})=V_s(\stackrel{\rightarrow}{q})
+\gamma_{_0}\otimes\gamma^0
V_v(\stackrel{\rightarrow}{q})~,\nonumber\\
&\displaystyle
V_s(\stackrel{\rightarrow}{q})=-(\frac{\lambda}{\alpha}+V_0)
\delta^3(\stackrel{\rightarrow}{q})+\frac{\lambda}{\pi^2}
\frac{1}{{(\stackrel{\rightarrow}{q}}^2+{\alpha}^2)^2}~,\nonumber\\
&\displaystyle
V_v(\stackrel{\rightarrow}{q})=-\frac{2}{3{\pi}^2}\frac{\alpha_s(
\stackrel{\rightarrow}{q})}{{(\stackrel{\rightarrow}{q}}^2+{\alpha}^2)}~\,,\nonumber\\
&\displaystyle
\alpha_s(\stackrel{\rightarrow}{q})=\frac{12\pi}{33-2N_f}\frac{1}
{\log
(a+\frac{{\stackrel{\rightarrow}{q}}^2}{\Lambda^{2}_{QCD}})}~\,
\end{eqnarray}
where $N_f=3$ for ($c \bar c$) system, $N_f=4$ for ($b \bar b$)
system; the constants $\lambda$, $\alpha$, $a$, $V_0$ and
$\Lambda_{QCD}$ are the parameters which characterize the kernel
(potential).

\section{General Formulation for the B.S. Wave Functions and the Coupled Equations}

In fact in this section, we show the new realized method to solve a
BS-In equation, but specifically apply to the concerned heavy
quarkonium problem.

Firstly, according to the total angle momentum ($J$), parity ($P$)
and charge conjugation ($C$) of the concerned bound state, we write
down the most general formulation for each of the relativistic B.S.
wave functions, and then we put it into Eq.(\ref{eq11}) to derive
out the coupled equation for the components appearing in the
formulation, BS-CoEq. In the below subsections, we do the derivation
for the low-laying states: $J^{PC}=0^{-+}, 1^{--}, 1^{+-}, 0^{++},
1^{++}, 2^{++}, etc$ in turn precisely.

\subsection{$J^{PC}=0^{-+},\, 1^{+-}, \, 0^{++}$ and $1^{++}$ wave functions
and BS-CoEqs for relevant components}

Since the bound states with the quantum numbers $J^{PC}=0^{-+},
1^{+-}, 0^{++}$ and $1^{++}$ are similar, so in this subsection we
derive the equations for them in turn.

I. {\it The bound states with quantum numbers $J^{PC}=0^{-+}$},
which in non-relativistic framework are $^1S_0$ states mainly.

The general formulation of the In-BS wave function Eq.(\ref{eq05})
for the states $J^{PC}=0^{-+}$ is \cite{cskimwang,changwang}:
\begin{equation}\label{eq10-+}
\varphi_{_{P,0^{-+}}}(q^\mu_{\perp})=\varphi_{_{P,0^{-+}}}(q_{_T})=
\left[\not\!Pf
_1(q_{_T})+Mf_2(q_{_T})+\frac{{\not\!P}{\not\!q}_{\perp}}{M}
f_3(q_{_T})\right]\gamma_{_5}\;,
\end{equation}
where $M$ is the mass of the bound state (the corresponding meson)
and $q^\mu_{\perp}=q^\mu-\frac{(P\cdot q)}{M^2}P^\mu$ is the four
dimensional vector. In the center mass system
$q^\mu_{\perp}=(0,\stackrel{\rightarrow}{q})$,
$q_{_T}=|\stackrel{\rightarrow}{q}|$.  Later on we abbreviate
$q_{_T}$ as $q$ if it does not make any confusion.

Now let us derive the coupled equations (BS-CoEqs) from
Eq.(\ref{eq11}). To put Eq.(\ref{eq10-+}) into the last two
equations of Eq.(\ref{eq11})
$$\varphi^{+-}_{_{P,0^{-+}}}(q)
=\varphi^{-+}_{_{P,0^{-+}}}(q)=0$$ and by taking various traces for
$\gamma$-matrices on both sides of the equations, we obtain the
independent constraints on the components for the wave function:
\begin{equation}\label{cons0-+}
{f_3(q)}=-\frac{f_1(q)M }{m_{1}}\;,
\end{equation}
so we can apply the obtained constraints Eq.(\ref{cons0-+}) to
Eq.(\ref{eq10-+}) and rewrite the relativistic wave function of
state $0^{-+}$ as:
\begin{eqnarray}\label{EQ:16}
\varphi_{_{P,0^{-+}}}(q)&=&\left[ {\not\!P}f_1(q) +Mf_2(q)+
{\not\!q}_{\perp}\frac{\not\!P}{m_1}
f_1(q)\right]\gamma_{_5}\nonumber\\
&=&\left[(1+\frac{\not\!q_{\perp}}{m_1}) {\not\!P}f_1(q)
+Mf_2(q)\right]\gamma_{_5}\;.
\end{eqnarray}
From the above formulation of the wave function one can see clearly
that besides the `great component', which is proportional to either
$M\gamma_5$ or ${\not\!P}\gamma_5$, there is also a `small
component', which is proportional to
$\frac{\not\!q_{\perp}}{m_1}{\not\!P}\gamma_5$, linear in
$q_{\perp}$ ($P$-wave nature) and suppressed by $\frac{1}{m_1}$.

Put the wave function Eq.(\ref{EQ:16}) into the first two equations
of Eq.(\ref{eq11}) and by taking various traces for
$\gamma$-matrices to both sides of the equations, we obtain the
independent coupled integral equations (BS-CoEqs):
\begin{eqnarray}\label{eq0-+}
&\displaystyle (M-2\omega_{1})\left[f_1({q})+
f_2({q})\frac{m_{1}}{\omega_{1}}\right]=
-\int\frac{d^3{\stackrel{\rightarrow}{k}}}{(2\pi)^3}
\frac{1}{\omega_{1}^2}\nonumber\\
&\displaystyle \times\left\{(V_s-V_v)\left[ f_1(k)m_{1}^2+
f_2(k)m_{1}\omega_{1} \right]-(V_s+V_v) f_1(k)
({\stackrel{\rightarrow}{q}}\cdot
{\stackrel{\rightarrow}{k}})\right\}~,\nonumber\\
&\displaystyle (M+2\omega_{1})\left[f_1({q})-
f_2({q})\frac{m_{1}}{\omega_{1}}\right]=
\int\frac{d^3{\stackrel{\rightarrow}{k}}}{(2\pi)^3}
\frac{1}{\omega_{1}^2}\nonumber\\
&\displaystyle \times\left\{(V_s-V_v)\left[ f_1(k)m_{1}^2-
f_2(k)m_{1}\omega_{1} \right]-(V_s+V_v) f_1(k)
({\stackrel{\rightarrow}{q}}\cdot
{\stackrel{\rightarrow}{k}})\right\}~,
\end{eqnarray}
here $k\equiv|\vec{k}|$, $\omega_{1}=\sqrt{m_{1}^{2}+q^{2}_{_T}}$.
Now we are prepare ready to solve the BS-CoEqs Eq.(\ref{eq0-+}), as
an eigenvalue problem, for $f_1$ and $f_2$ numerically, specially in
center mass system, and we may obtain the required results (the
spectrum for $J^{PC}=0^{-+}$ states and the B.S. wave functions
accordingly) finally.

Now accordingly the normalization condition is read as
\begin{equation}\int\frac{d^3{q}}{(2\pi)^3}4f_1
({q})f_2({q})M^2\left\{
\frac{\omega_{1}}{m_{1}}+\frac{m_{1}}{\omega_{1}}
+\frac{{q}^2}{\omega_{1}m_{1}} \right\}=2M~.
\end{equation}

II. {\it The bound states with quantum numbers $J^{PC}=1^{+-}$}
which in non-relativistic framework are $^1P_1$ states mainly:

As that for the states $J^{P}=1^{+-}$, the general form of the In-BS
wave function can be written as \cite{changwang,wang2}:
\begin{equation}\varphi_{_{P,1^{+-}}}(q)=
q_{\perp}\cdot{\epsilon}^{\lambda}_{\perp}\left[
f_1(q)+f_2(q)\frac{{\not\!P}}{M} +
f_3(q)\frac{\not\!{P}{\not\!q_{\perp}}}{M^2}\right]\gamma_5.
\end{equation}
From the equations
\begin{equation}
\varphi^{+-}_{_{P,1^{+-}}}(q)=\varphi^{-+}_{_{P,1^{+-}}}(q)=0\;,
\end{equation}
a constraint on the components of the wave function
$$f_3(q)=-\frac{f_2(q)M} {m_1}$$ is obtained.

With the constraint, the wave function now turns into:
\begin{eqnarray}\label{eq1+-0}
\varphi_{_{P,1^{+-}}}(q)&=&
q_{\perp}\cdot{\epsilon}^{\lambda}_{\perp}\left[
f_1(q)+f_2(q)\frac{{\not\!P}}{M} -f_2(q)
\frac{\not\!{P}{\not\!q_{\perp}}}{m_1M}\right]\gamma_5\nonumber\\
&=& q_{\perp}\cdot{\epsilon}^{\lambda}_{\perp}\left[
f_1(q)+f_2(q)\Big(1+\frac{\not\!q_{\perp}}{m_1}\Big)\frac{{\not\!P}}{M}
\right]\gamma_5\,,
\end{eqnarray}
here the factor ($q_{\perp}\cdot{\epsilon}^{\lambda}_{\perp}$)
indicates the wave function is of $P$-wave nature mainly; whereas
in Eq.(\ref{eq1+-0}) the `small component' term
$(q_{\perp}\cdot{\epsilon}^{\lambda}_{\perp})\frac{\not\!q_{\perp}}{m_1}
\frac{{\not\!P}}{M}\gamma_5$ contains high order wave.

Now the normalization condition for the $^1P_1$ wave function is
read as:
\begin{equation}
\int \frac{d^3{\vec q}}{(2\pi)^3}\frac{4f_1f_2
\omega_1{q}^2}{3m_1}=M.
\end{equation}

In terms of the same derivation as that for $J^{PC}=0^{-+}$ states,
we obtain the coupled equations (BS-CoEqs) for the components $f_1$
and $f_2$:
\begin{eqnarray}\label{eq1+-}
&\displaystyle
(M-2\omega_1)\left[f_1(q)+f_2(q)\frac{\omega_1}{m_1}\right]
=\int{\frac{d^3\vec{k}}{(2\pi)^3}\frac{(\vec{k}\cdot\vec{q})}
{\omega_1m_1q^2}}\nonumber\\
&\displaystyle \times\left\{(V_s+V_v)f_2(k)(\vec{k}\cdot \vec{q})
-(V_s-V_v)\left[f_1(k)\omega_1m_1
+f_2(k)m_1^2\right]\right\}\,,\nonumber\\
&\displaystyle
(M+2\omega_1)\left[f_1(q)-f_2(q)\frac{\omega_1}{m_1}\right]
=-\int{\frac{d^3\vec{k}}{(2\pi)^3}\frac{(\vec{k}\cdot\vec{q})}
{\omega_1m_1q^2}}\nonumber\\
&\displaystyle \times\left\{-(V_s+V_v)f_2(k)(\vec{k}\cdot \vec{q})
-(V_s-V_v)\left[f_1(k)m_1\omega_1 -f_2(k)m_1^2\right]\right\}\,.
\end{eqnarray}
Therefore, we are prepare ready to solve the coupled equations
BS-CoEqs, as an eigenvalue problem, for $f_1$ and $f_2$ numerically,
specially in center mass system, and we may obtain the required
results (the spectrum for $J^{PC}=1^{+-}$ states and the B.S. wave
function accordingly) finally.

III. {\it The bound states $J^{PC}=0^{++}$ and $1^{++}$} which in
non-relativistic framework essentially are $^3P_0$ and $^3P_1$
states respectively:

Since the bound states $J^{PC}=0^{++}$ and $1^{++}$ are very
similar, thus here we treat them simultaneously. For the states
$J^{P}=0^{++}$, the general form of the In-BS wave functions can be
written as \cite{changwang,wang2}:
\begin{equation}\varphi_{_{P,0^{++}}}(q)=
f_1(q){\not\!q_{\perp}}+f_2(q)\frac{{\not\!P} {\not\!q}_{\perp}}{M}
+f_3(q)M\;.
\end{equation}
With the equations
\begin{equation}
\varphi^{+-}_{_{P,0^{++}}}(q)=\varphi^{-+}_{_{P,0^{++}}}(q)=0\;,
\end{equation}
we obtain the constraints:
$$f_3(q)=-\frac{f_1(q)q^2}
{Mm_1}\;.$$ Then the wave function:
\begin{eqnarray}\label{eq0++10}
\varphi_{_{P,0^{++}}}(q)&=&
f_1(q){\not\!q_{\perp}}+f_2(q)\frac{{\not\!q_{\perp}}\not\!P}{M}
-\frac{f_1(q)\vec{q}~^2}{m_1}\nonumber\\
&=&{\not\!q_{\perp}}\Big[(1-\frac{\not\!q_{\perp}}{m_1})f_1(q)
+f_2(q)\frac{\not\!P}{M}\Big]\;.
\end{eqnarray}
Here the factor ${\not\!q_{\perp}}=(\vec{q}\cdot\vec{\gamma})$ in
CMS is contained in the wave functions, that means the the wave
function is of $P$-wave nature. Whereas the term which contains
$\vec{q}~^2=({\not\!q_{\perp}})({\not\!q_{\perp}})$ in
Eq.(\ref{eq0++10}) is suppressed by the factor $\frac{1}{M}$.

In terms of the same way as that for $J^{PC}=0^{-+}$ states, with
the first two equations of Eq.(\ref{eq11}) we obtain the coupled
equations for the $J^{PC}=0^{++}$ states:
\begin{eqnarray}\label{eq0++}
&\displaystyle(M-2\omega_1)\left[f_1(q)
+f_2(q)\frac{m_1}{\omega_1}\right]=\int{\frac{d^3\vec{k}}{(2\pi)^3}\frac{1}
{\omega_1^2}}
\times\left\{(V_s+V_v)\left[-f_1(k)q^2\right]k^2\right.\nonumber
\\
&+\left.(m_1)(V_s-V_v)\left[f_1(k)m_1+f_2(k)\omega_1\right](\vec{k}\cdot
\vec{q})\right\}\,;\nonumber\\
&\displaystyle
(M+2\omega_1)\left[f_1(q)-f_2(q)\frac{m_1}{\omega_1}\right]
=\int{\frac{d^3\vec{k}}{(2\pi)^3}\frac{1} {\omega_1^2}}
\times\left\{(V_s+V_v)\left[f_1(k)q^2\right]k^2\right.\nonumber\\
&-\left.(m_1)(V_s-V_v)\left[f_1(k)m_1-f_2(k)\omega_1\right](\vec{k}\cdot
\vec{q})\right\}\,.
\end{eqnarray}
The normalization condition for the wave function is read:
\begin{equation}
\int \frac{d^3{\vec q}}{(2\pi)^3}\frac{4f_1f_2
\omega_1{q^2}}{m_1}=M\,.
\end{equation}

Whereas for the $J^{PC}=1^{++}$ states, the general form for the
wave function can be written as \cite{changwang,wang2}:
\begin{eqnarray}
&\displaystyle
\varphi_{_{P,1^{++}}}(q)=i\varepsilon_{\mu\nu\alpha\beta}
P^{\nu}q_{\perp}^{\alpha}\epsilon^{\beta}\Big[f_1(q)M\gamma^{\mu}+f_2(q){\not\!P}\gamma^{\mu}
\nonumber\\
&\displaystyle +if_3(q)\varepsilon^{\mu\rho\sigma\delta}
q_{\perp\rho}P_{\sigma}\gamma_{\delta}\gamma_{5}/M \Big]/M^2\,.
\end{eqnarray}
From the he equations
\begin{equation}
\varphi^{+-}_{_{P,1^{++}}}(q)=\varphi^{-+}_{_{P,1^{++}}}(q)=0\,,
\end{equation}
we obtain the constraints on the components of the wave function:
$$f_3(q)=\frac{f_2(q)M}{m_1}\,.$$ Then we have:
\begin{equation}\label{eq1++0}
\varphi_{_{P,1^{++}}}(q)=i\varepsilon_{\mu\nu\alpha\beta}
P^{\nu}q_{\perp}^{\alpha}\epsilon^{\beta}\Big[f_1(q)M\gamma^{\mu}+
f_2(q)({\not\!P}\gamma^{\mu} +i\varepsilon^{\mu\rho\sigma\delta}
q_{\perp\rho}P_{\sigma}\gamma_{\delta}\gamma_{5}/m_1) \Big]/M^2\,.
\end{equation}
Here the front factor $\varepsilon_{\mu\nu\alpha\beta}
P^{\nu}q_{\perp}^{\alpha}\epsilon^{\beta}$, being linear in
$\vec{q}$, means the wave functions are of $P$-wave nature.

In terms of the same way as that for $J^{PC}=0^{-+}$ states, with
the first two equations of Eq.(\ref{eq11}) we obtain the coupled
equations (BS-CoEqs) as follows:
\begin{eqnarray}\label{eq1++}
&\displaystyle
(M-2\omega_1)\left[f_1(q)+f_2(q)\frac{\omega_1}{m_1}\right]
=\int{\frac{d^3\vec{k}}{(2\pi)^3}\frac{1}
{2\omega_1m_1\vec{k}^2\vec{q}^2}}
\times\left\{-(V_s+V_v)f_2(k)\left[k^2q^2+(\vec{k}\cdot
\vec{q})^2\right]\right.\nonumber \\
&-\left.2m_1(V_s-V_v)\left[f_1(k)\omega_1
+f_2(k)m_1\right]k^2(\vec{k}\cdot \vec{q})\right\}\nonumber\\
&\displaystyle(M+2\omega_1)\left[f_1(q)-f_2(q)\frac{\omega_1}{m_1}\right]
=-\int{\frac{d\vec{k}}{(2\pi)^3}\frac{1} {2\omega_1m_1k^2q^2}}
\times\left\{(V_s+V_v)f_2(k)\left[k^2q^2+(\vec{k}\cdot
\vec{q})^2\right]\right.\nonumber\\
&-\left.2m_1(V_s-V_v)\left[f_1(k)\omega_1
-f_2(k)m_1\right]k^2(\vec{k}\cdot \vec{q})\right\}
\end{eqnarray}

The normalization condition for the $J^{PC}=1^{++}$ wave function is
read:
\begin{equation}
\int \frac{d^3{\vec q}}{(2\pi)^3}\frac{8f_1f_2
\omega_1q^2}{3m_1}=M\,.
\end{equation}
Now we are ready to solve the coupled equations Eqs.(\ref{eq0++},
\ref{eq1++}) numerically.

\subsection{$J^{PC}=1^{--}, \,2^{++}$ wave functions for In-BS
equation and BS-CoEqs for relevant components}

As for the states $J^{PC}=1^{--}, 2^{++}$, they are quite different
from the states in the above subsection, because there is $S-D$ wave
mixing in the $J^{PC}=1^{--}$ states and there is $P-F$ wave mixing
in the $J^{PC}=2^{++}$ states.

I. {\it The bound states $J^{PC}=1^{--}$} which in non-relativistic
framework are $^3S_1$ and/or $^3D_1$ states mainly:

First of all, we write down the general formulation for the wave
functions of In-BS equation with quantum numbers
$J^P=1^{--}$\cite{changwang,wang1}:
\begin{eqnarray}
&\displaystyle\varphi_{_{P,1^{--}}}^{\lambda}(q)=
q_{\perp}\cdot{\epsilon}^{\lambda}_{\perp} \left[f_1(q)+
\frac{{\not\!q}_{\perp}}{M}f_3(q_{\perp})+\frac{{\not\!P}
{\not\!q}_{\perp}}{M^2} f_4(q)\right]+
M{\not\!\epsilon}^{\lambda}_{\perp}f_5(q)\nonumber\\
&\displaystyle+{\not\!\epsilon}^{\lambda}_{\perp}{\not\!P}f_6(q)
+\frac{1}{M}({\not\!P}{\not\!\epsilon}^{\lambda}_{\perp}
{\not\!q}_{\perp}-{\not\!P}q_{\perp}\cdot{\epsilon}^{\lambda}_{\perp})
f_2(q)\;,\label{eq013}
\end{eqnarray}
where the ${\epsilon}^{\lambda}_{\perp}$ is the polarization vector
of the vector meson. From the last two equations of Eq.(\ref{eq11})
\begin{equation}
\varphi^{\lambda,+-}_{_{P,1^{--}}}(q)=\varphi^{\lambda,-+}_{_{P,1^{--}}}(q)=0\;,
\end{equation}
we obtain the independent constraints on the components of the wave
functions:
$$f_1(q)=\frac{-q^2 f_3(q)
+M^2f_5(q)} {Mm_1}\,,~~~~f_2(q)=-\frac{f_6(q)M} {m_1}\,.$$ Then with
the constraints, there are only four independent components
$f_3(q)$, $f_4(q)$, $f_5(q)$ and $f_6(q)$ left in the
Eq.(\ref{eq013}). Namely
\begin{eqnarray}
&\displaystyle\varphi_{1^{--}}^{\lambda}(q_{\perp})=
q_{\perp}\cdot{\epsilon}^{\lambda}_{\perp}\Big(\frac{-q^2}{Mm_1}+
\frac{{\not\!q}_{\perp}}{M}\Big)f_3(q)+q_{\perp}\cdot{\epsilon}^{\lambda}_{\perp}
\frac{{\not\!P} {\not\!q}_{\perp}}{M^2} f_4(q) \nonumber\\
&+\displaystyle\Big( M{\not\!\epsilon}^{\lambda}_{\perp}
+q_{\perp}\cdot{\epsilon}^{\lambda}_{\perp}\frac{M}{m_1}\Big)f_5(q)
+\Big[{\not\!\epsilon}^{\lambda}_{\perp}{\not\!P}
+\frac{{\not\!P}(q_{\perp}\cdot{\epsilon}^{\lambda}_{\perp})}{m_1}
-\frac{({\not\!P}{\not\!\epsilon}^{\lambda}_{\perp}
{\not\!q}_{\perp})}{m_1}\Big] f_6(q), \label{eq1330}
\end{eqnarray}
and from the formulation it is easy to realize that the `great
components' in the wave function, which are proportional to  $f_5$
or $f_6$ and ${\not\!\epsilon}^{\lambda}_{\perp}$ or
$({\not\!\epsilon}^{\lambda}_{\perp}\not\!P)$ are of $S$-wave
nature, whereas the components in the wave function, which are
proportional to  $f_3$ or $f_4$ and
$(q_{\perp}\cdot{\epsilon}^{\lambda}_{\perp}){\not\!q}_{\perp}$ or
$(q_{\perp}\cdot{\epsilon}^{\lambda}_{\perp}){\not\!q}_{\perp}\not\!P$
(double $q_{\perp}$ being contained) are of $D$-wave nature (a
tensor about ${q}_{\perp}$). Therefore, no matter what are the other
`small terms', the wave functions Eq.(\ref{eq1330}) involve $S-D$
wave mixing properly.

To put Eq.(\ref{eq1330}) into the first two equations of
Eq.(\ref{eq11}) and take various traces on both sides of the
equations, we obtain four coupled integral equations for the four
independent components $f_3$, $f_4$, $f_5$ and $f_6$ (BS-CoEqs):
\begin{eqnarray}\label{eq1--01}
&\displaystyle
(M-2\omega_1)\left\{\left(f_3(q)\frac{q^2}{M^2}-f_5(q)\right)
+\left(f_4(q)\frac{q^2}{M^2}+f_6(q)\right)\frac{m_1}
{\omega_1}\right\}\nonumber\\
&\displaystyle=\int{\frac{d^3\vec{k}}{(2\pi)^3}\frac{2}{\omega_1^2}}\left\{(V_s+V_v)
\left(f_3(k)\frac{k^2}{M^2}-f_5(k)\right)
(\vec{k}\cdot\vec{q})\right.\nonumber\\
&\displaystyle-(V_s-V_v)\left[m_1^2\left(f_3(k)\frac{(\vec{k}\cdot
\vec{q})^2}{M^2q^2}-f_5(k)\right)\left.+m_1\omega_1\left(f_4(k)\frac{(\vec{k}\cdot
\vec{q})^2}{M^2q^2}+f_6(k)\right)\right]\right\}\,,
\end{eqnarray}
$$(M+2\omega_1)\left\{\left(f_3(q)\frac{q^2}{M^2}-f_5(q)\right)
-\left(f_4(q)\frac{q^2}{M^2}+f_6(q)\right)\frac{m_1}
{\omega_1}\right\}$$
$$=-\int{\frac{d^3\vec{k}}{(2\pi)^3}\frac{2}{\omega_1^2}}\left\{(V_s+V_v)\left[
\left(f_3(k)\frac{k^2}{M^2}-f_5(k)\right)
\right](\vec{k}\cdot\vec{q})\right.$$
\begin{equation}\label{eq1--02}
-(V_s-V_v)\left[m_1^2\left(f_3(k)\frac{(\vec{k}\cdot
\vec{q})^2}{M^2q^2}-f_5(k)\right)\left.-m_1\omega_1\left(f_4(k)\frac{(\vec{k}\cdot
\vec{q})^2}{M^2q^2}+f_6(k)\right)\right]\right\}\,,
\end{equation}
$$(M-2\omega_1)\left\{\left(f_3(q)+f_4(q)\frac{m_1}{\omega_1}\right)
\frac{q^2}{M^2} -3\left(f_5(q)-f_6(q)\frac
{\omega_1}{m_1}\right)-f_6(q)\frac{q^2}{m_1\omega_1}\right\}$$
$$
=-\int{\frac{d^3\vec{k}}{(2\pi)^3}\frac{1}{\omega_1^2}}\left\{(V_s+V_v)
\left[-\frac{2\omega_1}{m_1} f_6(k)-f_3(k)\frac{k^2}{M^2}+
f_5(k)\right](\vec{k}\cdot \vec{q})\right.$$
$$
+(V_s-V_v)\left[\omega_1^2\left(f_3(k)\frac{k^2}{M^2}
-3f_5(k)\right)+m_1\omega_1
\left(f_4(k)\frac{k^2}{M^2}+3f_6(k)\right)\right.$$
\begin{equation}\label{eq1--03}
\left.\left. -\left(f_3(k)\frac{(\vec{k}\cdot\vec{q})^2}{M^2}
-f_5(k)\vec{q}^2\right)\right]\right\}\,,
\end{equation}
$$(M+2\omega_1)\left\{\left[f_3(q)-f_4(q)\frac{m_1}{\omega_1}\right]
\frac{q^2}{M^2}-3\left(f_5(q)+f_6(q)\frac
{\omega_1}{m_1}\right)+f_6(q)\frac{q^2}{m_1\omega_1}\right\}$$
$$
=\int{\frac{d^3\vec{k}}{(2\pi)^3}\frac{1}{\omega_1^2}}\left\{(V_s+V_v)
\left[\frac{2\omega_1}{m_1}f_6(k) -f_3(k)\frac{k^2}{M^2}+
f_5(k)\right](\vec{k}\cdot\vec{q})\right.$$
$$
+(V_s-V_v)\left[\omega_1^2\left(f_3(k)\frac{k^2}{M^2}-3f_5(k)\right)-m_1\omega_1
\left(f_4(k)\frac{k^2}{M^2}+3f_6(k)\right)\right.$$
\begin{equation}\label{eq1--04}
\left.\left. -\left(f_3(k)\frac{(\vec{k}\cdot\vec{q})^2}{M^2}
-f_5(k)q^2\right)\right]\right\}\,.
\end{equation}

Now the normalization condition for the wave functions with the
components $f_3({q})$, $f_4({q})$, $f_5({q})$ and $f_6({q})$ is read
as follows:
\begin{equation}
\int \frac{d{\vec q}}{(2\pi)^3}\frac{16\omega_1\omega_2}{3}\left\{
3f_5f_6\frac{M^2}{2m_1\omega_1}+\frac{q^2}{2m_1\omega_1}\left[
f_4f_5-f_3\left(f_4\frac{ q^2}{M^2}+f_6\right)\right] \right\}=2M.
\end{equation}

Thus the results about the mass spectra and the wave functions for
the $J^{PC}=1^{--}$ bound states can be obtained by solving the
coupled Eqs.(\ref{eq1--01}-\ref{eq1--04}) numerically.

II. {\it The bound states with quantum numbers $J^{PC}=2^{++}$}
which in non-relativistic framework are $^3P_2$ and/or $^3F_2$
states mainly:

The general form of the wave function for $J^{PC}=2^{++}$ states can
be written down as:

\begin{eqnarray}
\varphi_{_{P,2^{++}}}^{\lambda}(q)=
{\varepsilon}^{\lambda}_{\mu\nu}{q_{\perp}^{\nu}}
\Bigg\{{q_{\perp}^{\mu}}\Big[f_1(q)+
\frac{{\not\!q}_{\perp}}{M}f_3(q)+\frac{{\not\!P}
{\not\!q}_{\perp}}{M^2} f_4(q)\Big]\nonumber\\
+ {\gamma^{\mu}}\Big[Mf_5(q)+ {\not\!P}f_6(q)\Big]+\frac{i}{M}
f_2(q)\epsilon^{\mu\alpha\beta\gamma}
P_{\alpha}q_{\perp\beta}\gamma_{\gamma}\gamma_{5}\Bigg\},\label{eq233}
\end{eqnarray}
where the ${\varepsilon}^{\lambda}_{\mu\nu}$ (symmetric
in $\mu$ and $\nu$) is the tensor polarization of the meson. From
the last two equations of Eq.(\ref{eq11}):
\begin{equation}
\varphi^{\lambda,+-}_{_{P,2^{++}}}(q)=\varphi^{\lambda,-+}_{_{P,2^{++}}}(q)=0\,,
\end{equation}
we obtain the constraints on the components of the wave function:
$$f_1(q)=\frac{-q^2 f_3(q)+M^2f_5(q)
} {Mm_1}\,,~~~~~f_2(q)=\frac{f_6(q)M} {m_1}\,.$$ Put the constraints
into Eq.(\ref{eq233}), then only four independent components
$f_3(q)$\, $f_4(q)$, $f_5(q)$ and $f_6(q)$ are left:
\begin{eqnarray}
&\displaystyle\varphi_{_{P,2^{++}}}^{\lambda}(q_{\perp})=
{\varepsilon}^{\lambda}_{\mu\nu}{q_{\perp}^{\nu}}
\left\{{q_{\perp}^{\mu}}\left[\Big(
\frac{{\not\!q}_{\perp}}{M}-\frac{q^2}{Mm_1}\Big)f_3(q)+\frac{{\not\!P}
{\not\!q}_{\perp}}{M^2} f_4(q)\right]\right.\nonumber \\
&\displaystyle \left.+
({\gamma^{\mu}}+\frac{q_{\perp}^{\mu}}{m_1})Mf_5(q)+
{\gamma^{\mu}}{\not\!P}f_6(q)+i\frac{f_6(q)}
{m_1}\epsilon^{\mu\alpha\beta\gamma}
P_{\alpha}q_{\perp\beta}\gamma_{\gamma}\gamma_{5}\right\}\,.\label{eq2330}
\end{eqnarray}
From the formulation Eq.(\ref{eq2330}), it is easy to see that the
terms, which is proportional to $f_5$ or $f_6$ and with the factor
$({\varepsilon}^{\lambda}_{\mu\nu}{q_{\perp}^{\nu}}\gamma^\mu) M$ or
$({\varepsilon}^{\lambda}_{\mu\nu}{q_{\perp}^{\nu}}\gamma^\mu)\not\!P$,
are of $P$-wave nature (linear in ${q}_{\perp}$), and the terms,
which proportional to $f_3$ or $f_4$ and
$({\varepsilon}^{\lambda}_{\mu\nu}{q_{\perp}^{\nu}}{q_{\perp}^{\mu}})
\frac{\not\!P}{M}$ or
$({\varepsilon}^{\lambda}_{\mu\nu}{q_{\perp}^{\nu}}{q_{\perp}^{\mu}})
\frac{\not\!P\not\!q_{\perp}}{M^2}$, are of $F$-wave nature (cubic
in ${q}_{\perp}$). Therefore, Eq.(\ref{eq2330}) describes $P-F$ wave
mixing properly.

Put Eq.(\ref{eq2330}) into the first two equations of
Eq.(\ref{eq11}) and take various traces for $\gamma$-matrix on both
sides of the equations, we obtain the coupled equations for the four
independent components $f_3$, $f_4$, $f_5$ and $f_6$ as follows:
\begin{eqnarray}\label{eq2++01}
&\displaystyle
(M-2\omega_1)\left\{\left(f_3({q})\frac{{q}^2}{M^2}-f_5({q})\right)
+\left(f_4({q})\frac{{q}^2}{M^2}+f_6({q})\right)
\frac{m_1}{\omega_1}\right\}\nonumber \\
&\displaystyle=\int{\frac{d^3\vec{k}}{(2\pi)^3}\frac{1}{2\omega_1^2{q}^4}}\left\{-(V_s+V_v)
{q}^2\left(f_3({k})\frac{{k}^2}{M^2}-f_5({k})\right)
\left[{k}^2{q}^2-3(\vec{k}\cdot\vec{q})^2\right]
\right.\nonumber\\
&\displaystyle+m_1(V_s-V_v)\left[m_1\left(f_3({k})\frac{{k}^2{q}^2-3(\vec{k}\cdot\vec{q})^2}
{M^2}+2f_5(\vec{k}){q}^2\right)\right. \nonumber\\
&\displaystyle\left.\left.
+\omega_1\left(f_4({k})\frac{{k}^2{q}^2-3(\vec{k}\cdot\vec{q})^2}{M^2}
-2f_6({k}){q}^2\right)\right] (\vec{k}\cdot\vec{q})\right\}\,;
\end{eqnarray}
\begin{eqnarray}\label{eq2++02}
&\displaystyle
(M+2\omega_1)\left\{\left(f_3({q})\frac{{q}^2}{M^2}-f_5({q})\right)
-\left(f_4({q})\frac{{q}^2}{M^2}+f_6({q})\right)
\frac{m_1}{\omega_1}\right\}\nonumber\\
&\displaystyle=-\int{\frac{d^3\vec{k}}{(2\pi)^3}\frac{1}{2\omega_1^2{q}^4}}\left\{-(V_s+V_v)
{q}^2\left(f_3({k})\frac{{k}^2}{M^2}-f_5({k})\right)
\left[{k}^2{q}^2-3(\vec{k}\cdot\vec{q})^2\right]
\right.\nonumber\\
&\displaystyle+m_1(V_s-V_v)\left[ m_1\left(f_3({k})\frac{{k}^2{q}^2
-3(\vec{k}\cdot\vec{q})^2}{M^2}+2f_5({k}){q}^2\right)\right.\nonumber\\
&\displaystyle\left.\left.-\omega_1\left(f_4({k})\frac{{k}^2{q}^2-3(\vec{k}\cdot\vec{q})^2}{M^2}
-2f_6({k}){q}^2\right) \right] (\vec{k}\cdot\vec{q})\right\}\,;
\end{eqnarray}
\begin{eqnarray}\label{eq2++03}
&\displaystyle(M-2\omega_1)\left\{-f_5({q})m_1+f_6({q})
\omega_1\right\}=\int{\frac{d^3\vec{k}}{(2\pi)^3}\frac{1}
{\omega_1{q}^2}}\nonumber\\
&\displaystyle \times\left\{-\frac{1}{2}(V_s+V_v)f_6({k})
\left[{k}^2{q}^2-3(\vec{k}\cdot\vec{q})^2\right]+(V_s-V_v)m_1
\left[\left(f_5({k})\omega_1-f_6({k})m_1\right)\right.\right.\nonumber\\
&\displaystyle \left.\left.-\left(f_3({k})
\omega_1+f_4({k})m_1\right)\frac{{k}^2}{M^2}+
\left(f_3({k})\omega_1+f_4({k})m_1\right)\frac{(\vec{k}\cdot\vec{q})^2}{M^2{q}^2}\right]
(\vec{k}\cdot\vec{q})\right\}\,;
\end{eqnarray}
\begin{eqnarray}\label{eq2++04}
&\displaystyle(M+2\omega_1)\left\{-f_5({q})m_1-f_6({q})\omega_1\right\}
=-\int{\frac{d^3\vec{k}}{(2\pi)^3}\frac{1}
{\omega_1{q}^2}}\nonumber\\
&\displaystyle\times\left\{\frac{1}{2}(V_s+V_v)f_6({k})
\left[{k}^2{q}^2-3(\vec{k}\cdot\vec{q})^2\right]
-(V_s-V_v)m_1
\left[\left(f_5({k})\omega_1+f_6({k})m_1\right)\right.\right.\nonumber\\
&\displaystyle \left.\left.-\left(f_3({k})
\omega_1-f_4({k})m_1\right)\frac{{k}^2}{M^2}+\left(f_3({k})
\omega_1-f_4({k})m_1\right)\frac{(\vec{k}\cdot\vec{q})^2}{M^2{q}^2}\right]
(\vec{k}\cdot\vec{q})\right\}
\end{eqnarray}

Now the normalization condition is read as:
\begin{equation}
\int\frac{d^3{\vec q}}{(2\pi)^3}\frac{8 \omega_1 q^2}{15m_1}\left\{
5f_5f_6M^2 +2f_4f_5 {q}^2 -2
q^2f_3\left(f_4\frac{q^2}{M^2}+f_6\right) \right\}=2M.
\end{equation}

By solving the coupled equations Eqs.(\ref{eq2++01}-\ref{eq2++04})
numerically, we obtain the mass spectra and relevant B.S. wave
functions for the $J^{PC}=2^{++}$ bound states.

In fact, with the way described here, one may derive the BS-In
equation for the other possible $J^{PC}$ states into their BS-CoEqs
according one's wish.

\section{Numerical Results and Discussions}

\begin{table}[]\begin{center}
\caption{Parameter of $V_0$ in unit of $MeV$.} \vspace{0.5cm}
\begin{tabular}
{|c|c|c|}\hline\hline

&$c\bar c$&$b\bar b$ \\\hline\hline

{\bf n}$J^{PC}={\bf n}~0^{-+}(^1S_0)$&-0.314&-0.240 \\\hline

{\bf n}$J^{PC}={\bf n}~1^{--}(^3S_1)$&-0.176&-0.166 \\\hline\hline

{\bf n}$J^{PC}={\bf n}~0^{++}(^3P_0)$&-0.282&-0.174 \\\hline

{\bf n}$J^{PC}={\bf n}~1^{++}(^3P_1)$&-0.162&-0.141 \\\hline

{\bf n}$J^{PC}={\bf n}~2^{++}(^3P_2)$&-0.110&-0.121 \\\hline\hline

{\bf n}$J^{PC}={\bf n}~1^{+-}(^1P_1)$&-0.144&-0.135 \\\hline

\end{tabular}
\end{center}
\end{table}

\begin{table}[]\begin{center}
\caption{Mass spectra of $(c\bar c)$ and $(b\bar b)$ systems with
quantum numbers $J^{PC}=0^{-+}$, $1^{--}$. Here $^{(2S+1)}L_J$
denotes the dominant component in the state respectively. `Ex' means
the experimental results from PDG \cite{PDG} (and the data for
$\eta_b$ come from reference \cite{etab}.} \vspace{0.5cm}
\begin{tabular}
{|c|c|c|c|c|c|c|}\hline

{\bf n}~$ J^{PC} (^{(2S+1)}L_J)$&Th($c\bar c$)&Ex($c\bar
c$)&Th($b\bar b$)&Ex($b\bar b$)
\\\hline\hline

{\bf 1}$~0^{-+}(^1S_0)$&2980.3(input)&2980.3&9390.2(input)&9388.9
\\\hline

{\bf 2}$~0^{-+}(^1S_0)$&3576.4&3637&9950.0& \\\hline

{\bf 3}$~0^{-+}(^1S_0)$&3948.8&&10311.4&
\\\hline\hline

{\bf 1}$~1^{--}(^3S_1)$&3096.9(input)&3096.916&9460.5(input)&9460.30
\\\hline

{\bf 2}$~1^{--}(^3S_1)$&3688.1&3686.09&10023.1&10023.26
\\\hline

{\bf 3}$~1^{--}(^3D_1)$&3778.9&3772.92&10129.5&
\\\hline

{\bf 4}$~1^{--}(^3S_1)$&4056.8&4039&10368.9&10355.2
\\\hline

{\bf 5}$~1^{--}(^3D_1)$&4110.7&4153&10434.7& \\\hline

{\bf 6}$~1^{--}(^3S_1)$&4329.4&4421&10635.8&10579.4
\\\hline

{\bf 7}$~1^{--}(^3S_1)$&4545.9&&10852.1&10865 \\\hline

\end{tabular}
\end{center}
\end{table}

\begin{table}[]\begin{center}
\caption{Mass spectra of $(c\bar c)$ and $(b\bar b)$ systems with
quantum numbers $J^{PC}=0^{++}$, $1^{++}$, $2^{++}$, $1^{+-}$ in
unit of $MeV$. Here $^{(2S+1)}L_J$ denotes the dominant component in
the state respectively. `Ex' means the experimental results from PDG
\cite{PDG}.}

\vspace{0.5cm}
\begin{tabular}
{|c|c|c|c|c|c|}\hline

{\bf n}~$J^{PC}(^{2S+1)}L_J$&Th($c\bar c$)&Ex($c\bar c$)&Th($b\bar
b$)&Ex($b\bar b$)
\\\hline\hline

{\bf 1}$~0^{++}(^3P_0)$& 3414.7(input)&3414.75&9859.0&9859.44
\\\hline

{\bf 2}$~0^{++}(^3P_0)$& 3836.8 & &10240.6&10232.5
\\\hline

{\bf 3}$~0^{++}(^3P_0)$&4140.1&&10524.7&
\\\hline\hline

{\bf 1}$~1^{++}(^3P_1)$&3510.3(input)&3510.66&9892.2&9892.78
\\\hline

{\bf 2}$~1^{++}(^3P_1)$&3928.7&&10272.7&10255.46
\\\hline

{\bf 3}$~1^{++}(^3P_1)$&4228.8&&10556.2&
\\\hline\hline

{\bf 1}$~2^{++}(^3P_2)$&3556.1(input)&3556.20&9914.4&9912.21
\\\hline

{\bf 2}$~2^{++}(^3P_2)$&3972.4&&10293.6&10268.65
\\\hline

{\bf 3}$~2^{++}(^3F_2)$&4037.9&&10374.4& \\\hline

{\bf 4}$~2^{++}(^3P_2)$&4271.0&&10561.5&
\\\hline\hline

{\bf 1}$~1^{+-}(^1P_1)$&3526.0(input)&3525.93&9900.2&
\\\hline

{\bf 2}$~1^{+-}(^1P_1)$&3943.0&&10280.4& \\\hline

{\bf 3}$~1^{+-}(^1P_1)$&4242.4&&10562.0& \\\hline
\end{tabular}
\end{center}
\end{table}

\begin{figure}
\centering
\includegraphics[width=0.43\textwidth]{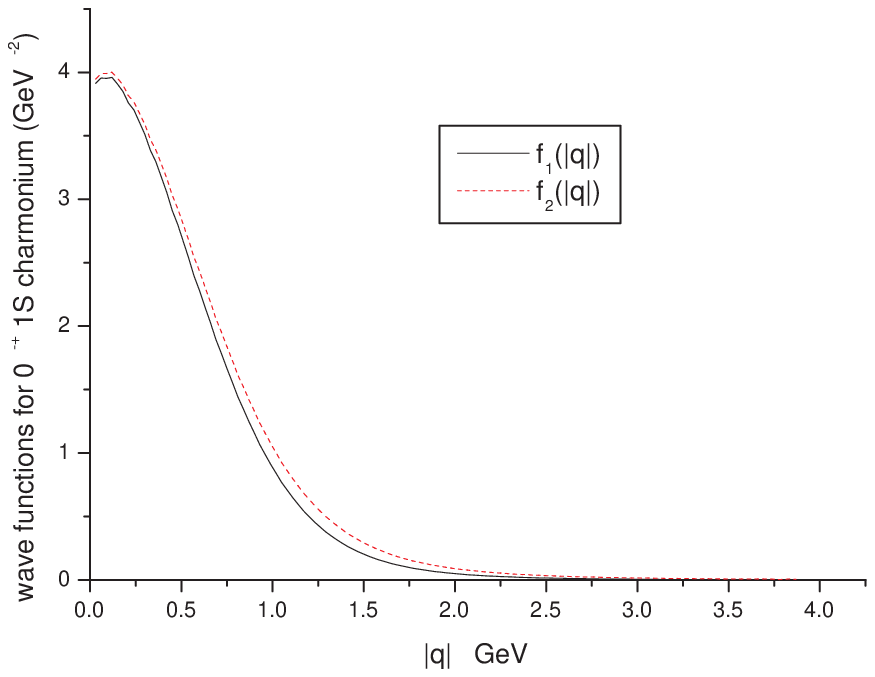}
\includegraphics[width=0.43\textwidth]{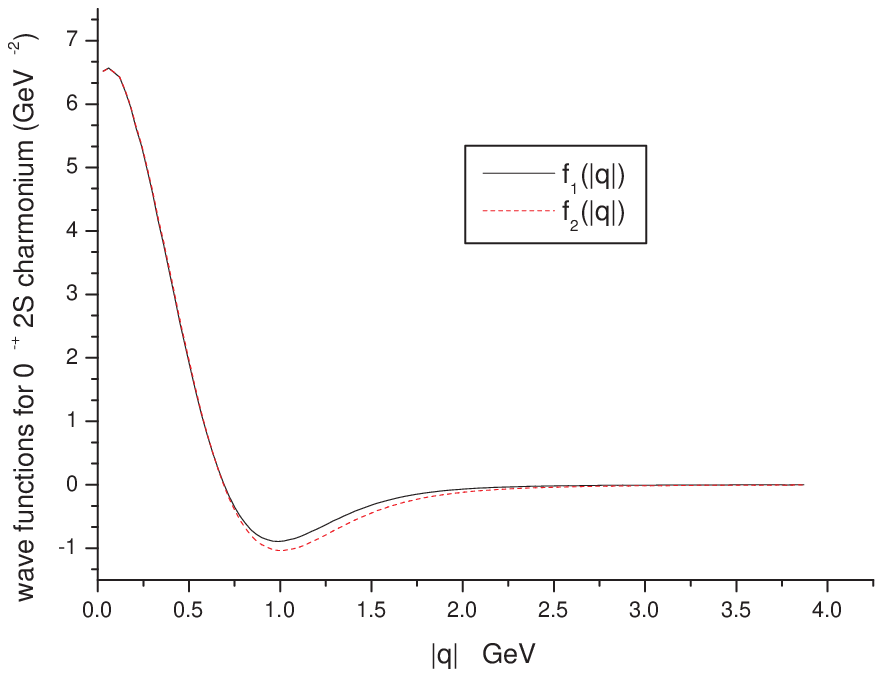}
\includegraphics[width=0.43\textwidth]{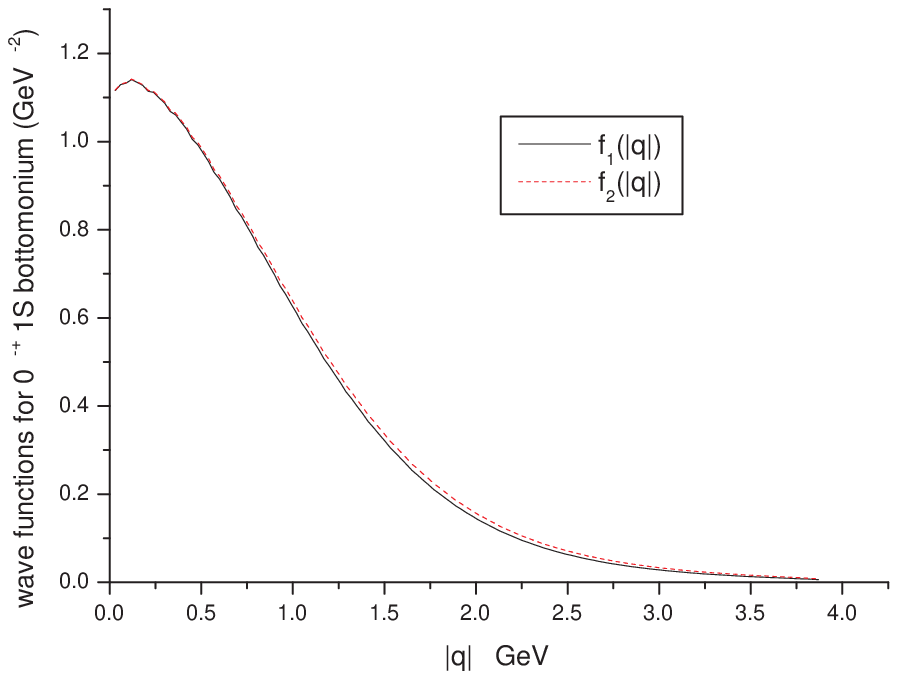}
\includegraphics[width=0.43\textwidth]{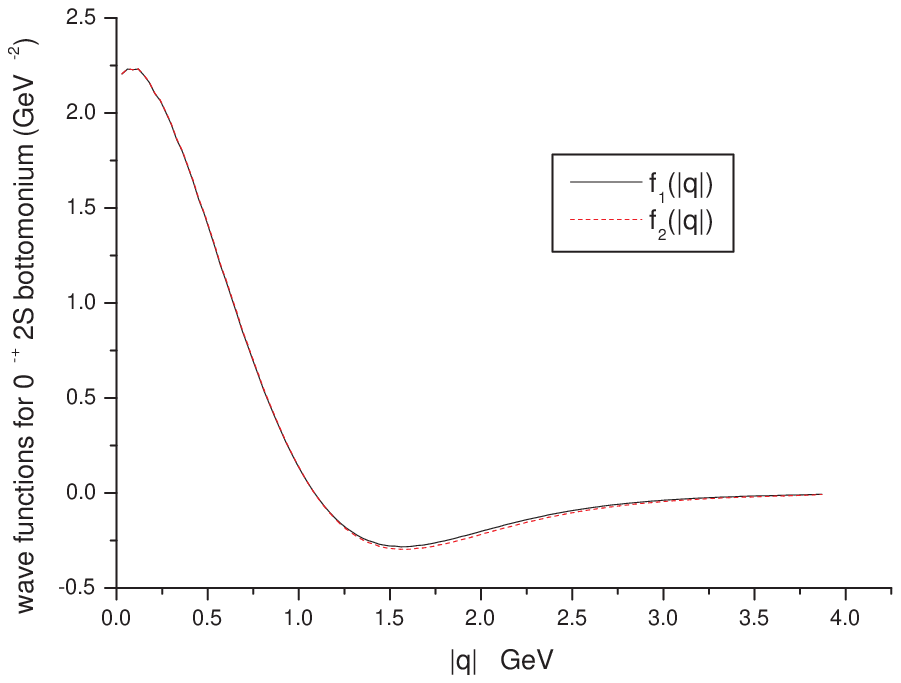}
\caption{The solutions for the wave functions of the ground and the
first excited state (from left to right) with quantum number
$J^{PC}=0^{-+}$. The wave functions (solutions) of the low-lying
states (the ground and the first excited state) with quantum number
$J^{PC}=0^{-+}$. The above two are those for charmonium and the
below two are those for bottonium.} \label{pdft1}
\end{figure}

\begin{figure}
\centering
\includegraphics[width=0.43\textwidth]{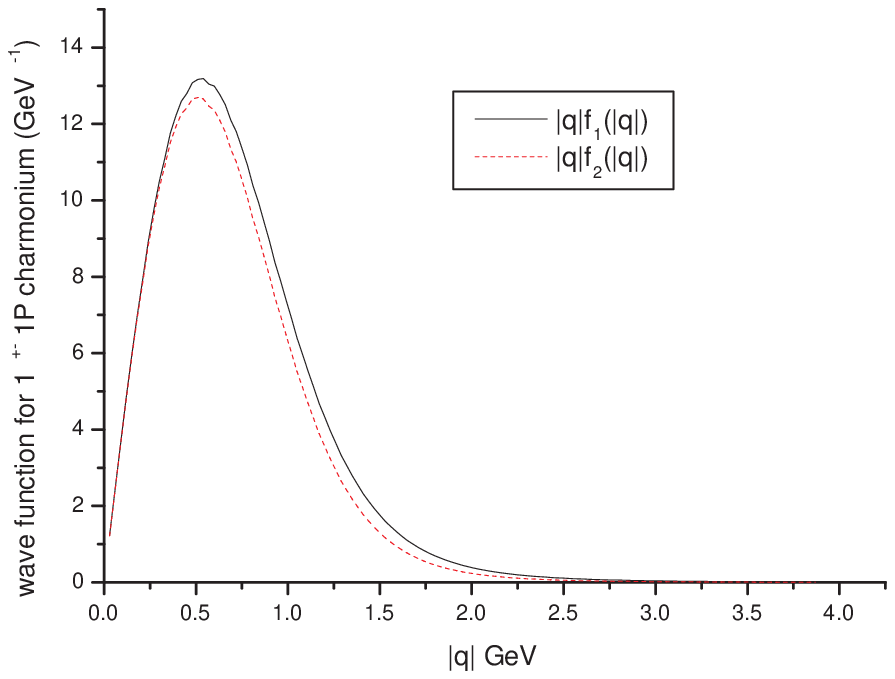}
\includegraphics[width=0.43\textwidth]{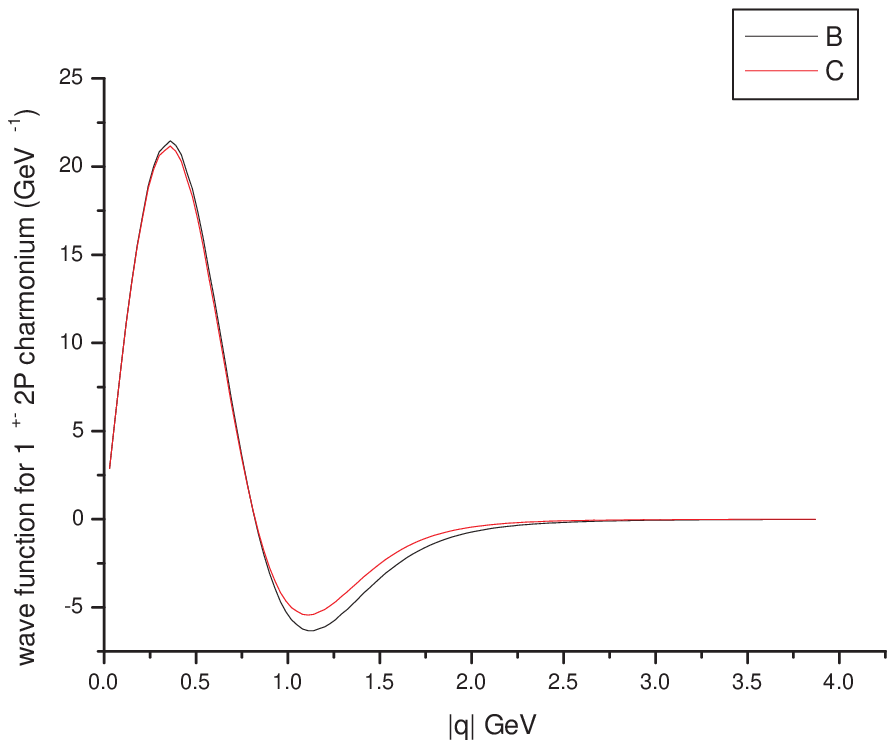}
\includegraphics[width=0.43\textwidth]{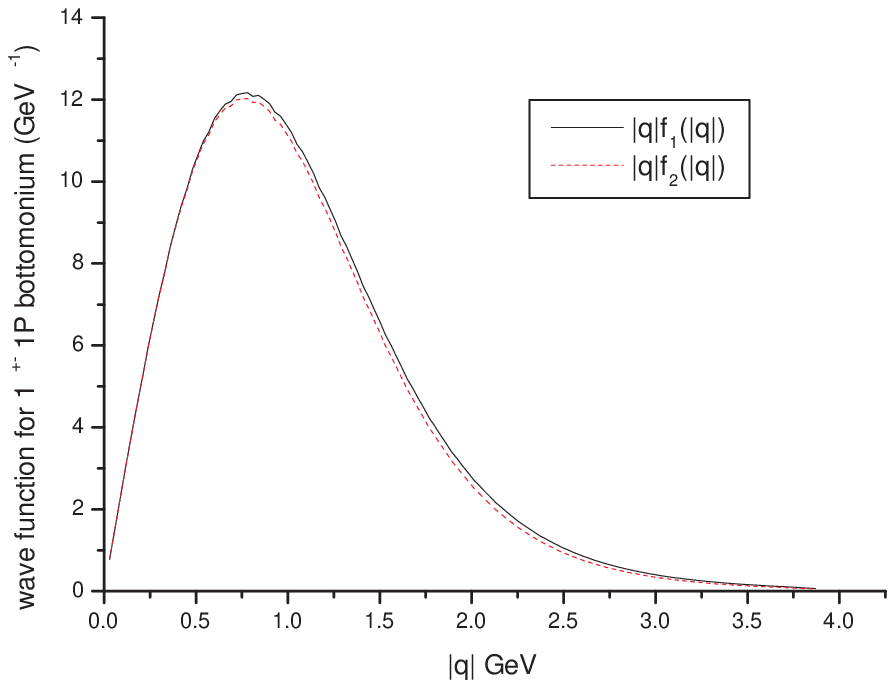}
\includegraphics[width=0.43\textwidth]{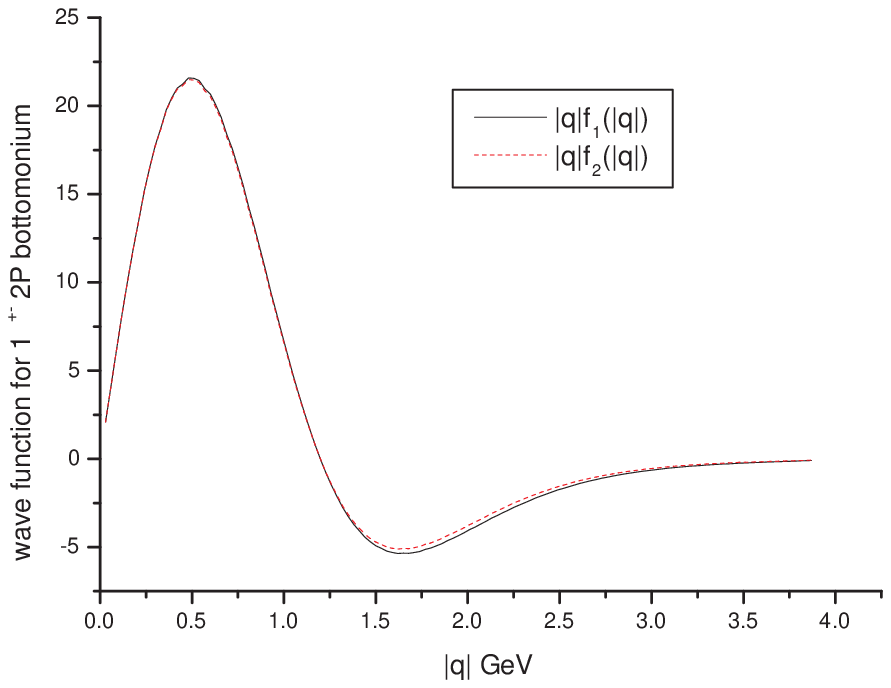}
\caption{The solutions for the wave functions of the ground and the
first excited state (from left to right) with quantum number
$J^{PC}=1^{+-}$. The wave functions (solutions) of the low-lying
states (the ground and the first excited state) with quantum number
$J^{PC}=1^{+-}$. The above two are those for charmonium and the
below two are those for bottonium.} \label{pdft2}
\end{figure}

\begin{figure}
\centering
\includegraphics[width=0.43\textwidth]{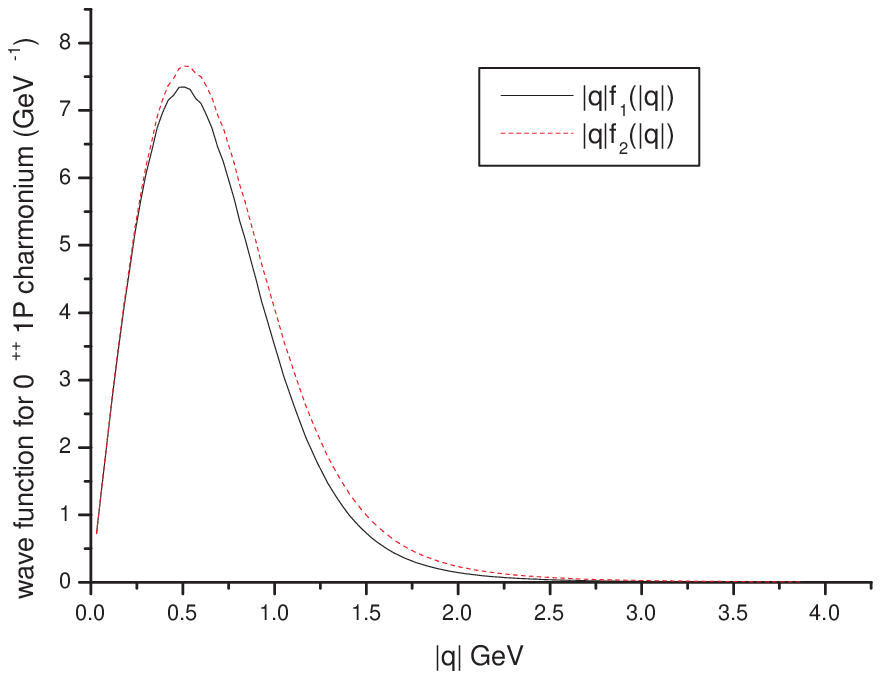}
\includegraphics[width=0.43\textwidth]{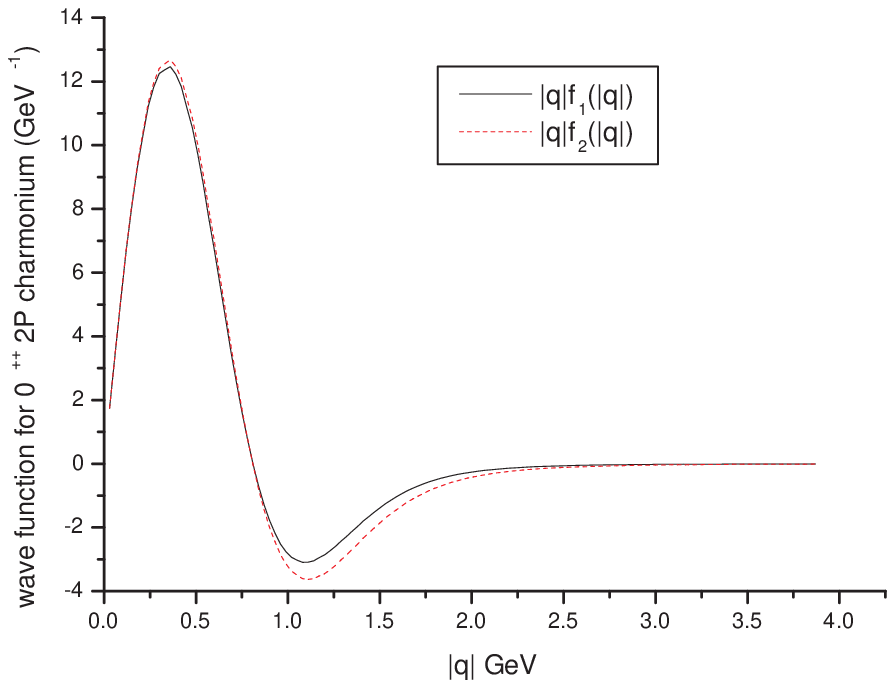}
\includegraphics[width=0.43\textwidth]{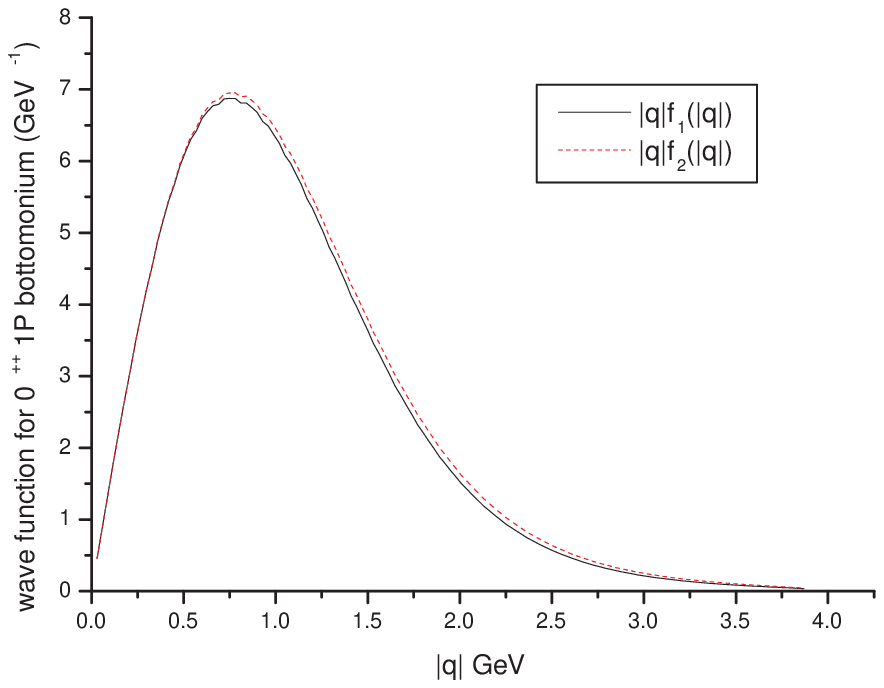}
\includegraphics[width=0.43\textwidth]{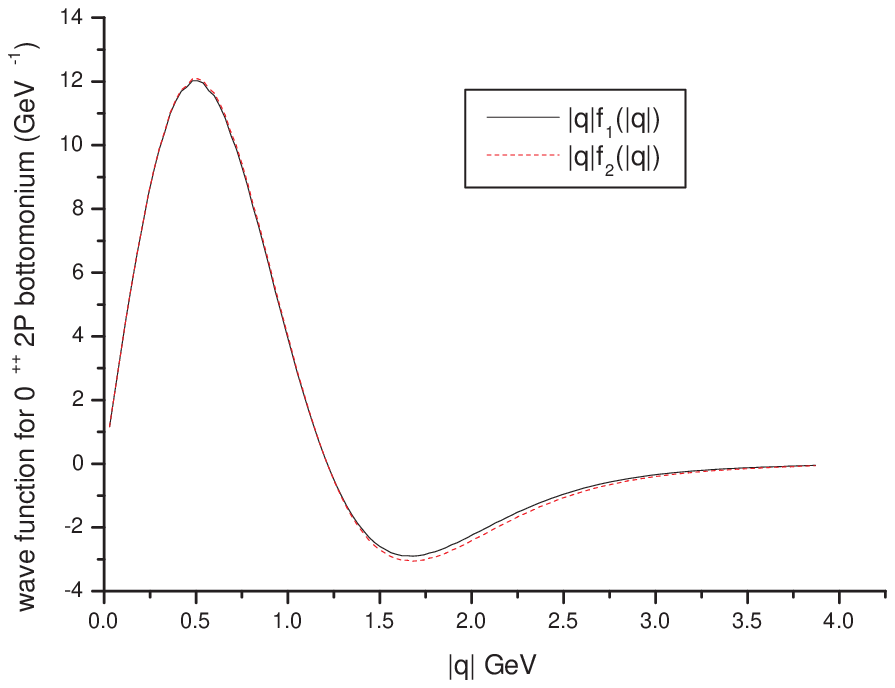}
\caption{The solutions for the wave functions of the ground and the
first excited state (from left to right) with quantum number
$J^{PC}=0^{++}$. The wave functions (solutions) of the low-lying
states (the ground and the first excited state) with quantum number
$J^{PC}=0^{++}$. The above two are those for charmonium and the
below two are those for bottonium.} \label{pdft3}
\end{figure}

\begin{figure}
\centering
\includegraphics[width=0.43\textwidth]{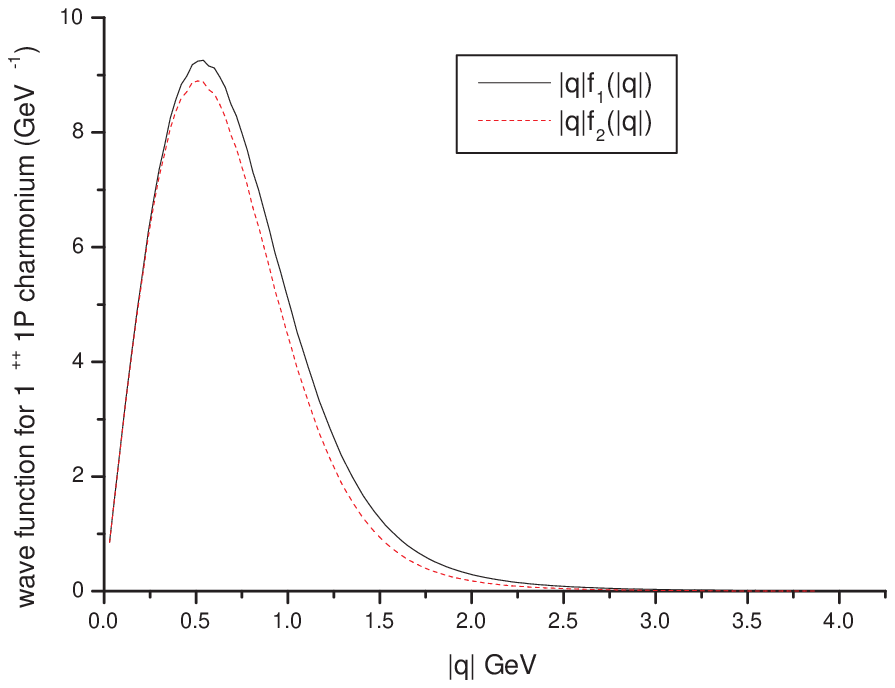}
\includegraphics[width=0.43\textwidth]{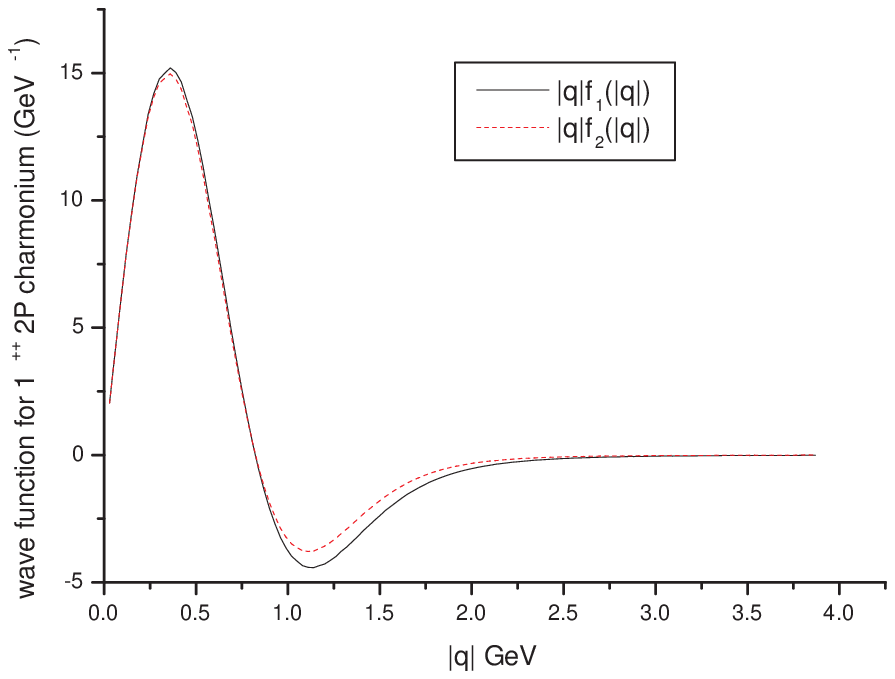}
\includegraphics[width=0.43\textwidth]{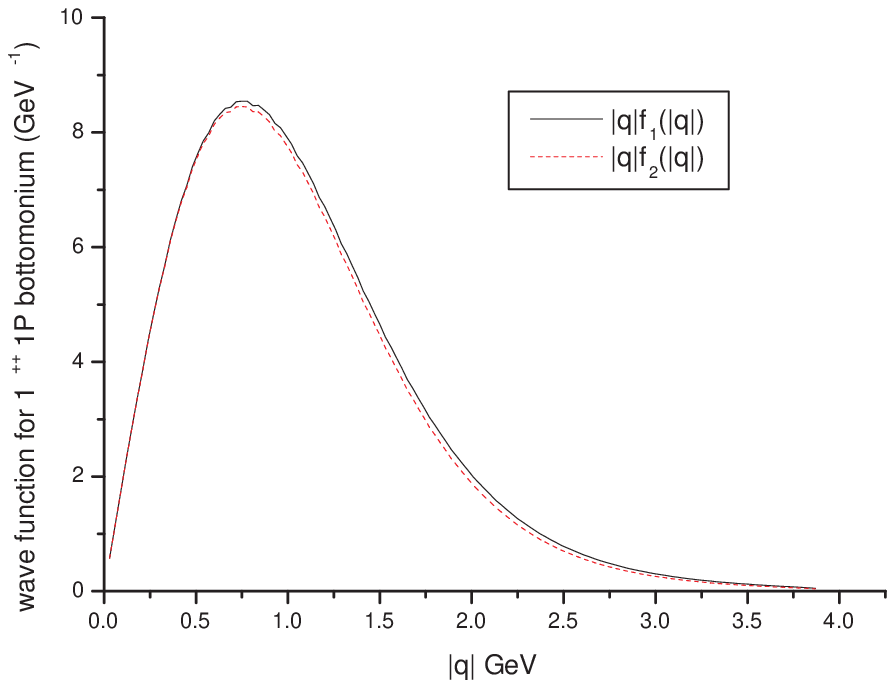}
\includegraphics[width=0.43\textwidth]{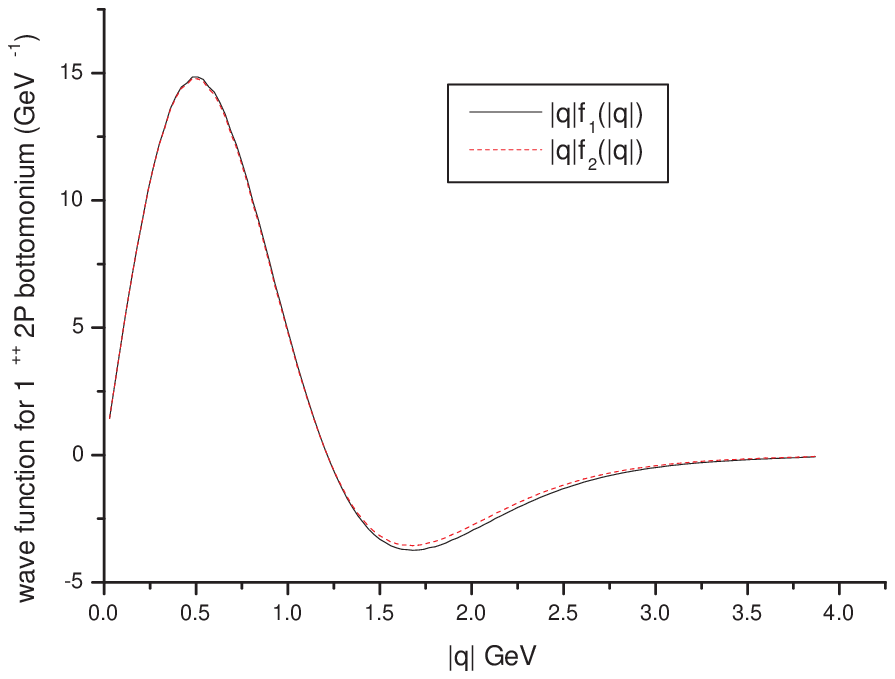}
\caption{The solutions for the wave functions of the ground and the
first excited state (from left to right) with quantum number
$J^{PC}=1^{++}$. The wave functions (solutions) of the low-lying
states (the ground and the first excited state) with quantum number
$J^{PC}=1^{++}$. The above two are those for charmonium and the
below two are those for bottonium.} \label{pdft4}
\end{figure}

In this section we solve the equations (BS-CoEqs) numerically and
discuss the obtained results.

Since the coupled integration equations are quite complicated, so we
solve them numerically only, and additionally with certain
approximation such as that a cut on the up-bound of the integrations
in the equations has been made.

To solve the equations, we also need to fix the parameters appearing
in the kernel Eq.(\ref{eq-poten}) although the kernel is based on
QCD inspirer and the Cornell potential for non-relativistic heavy
quark model as reference. Usually, the parameters are fixed by
fitting the best experimental data. Since now quite a lot of data
about the charmonium and bottomnium with quantum data
$J^{PC}=1^{--}$ are available and quite precise, so the most
parameters are fixed by the data. Since $V_0$ in the kernel
originates from QCD non-perturbative effects, its value is to
account the states with various $J^{PC}$, so we fix it by fitting
the mass of the ground states. Thus the parameter $V_0$ vary with
$J^{PC}$.

By fitting data, the values of the parameters for all of the states
are those as follows:
\begin{eqnarray}
&a=e=2.7183~, ~~\alpha=0.06~ {\rm GeV}, ~~\lambda=0.21~ {\rm
GeV}^2,\nonumber\\
&{\rm and}~~ m_c=1.62~ {\rm GeV},~~m_b=4.96~ {\rm GeV}\,.
\label{para}
\end{eqnarray}

Since the running coupling constant is used, so we also need to fix
$~\Lambda_{QCD}$. There are three active flavors for ($c \bar c$)
system, i.e. $N_f=3$, accordingly we adopt $~\Lambda_{QCD}=0.27~
{\rm GeV}$ and the coupling constant at the scale of charm quark
mass, $\alpha_s(m_c)=0.38$. There are four active flavors for ($b
\bar b$) system, i.e. $N_f=4$, so $~\Lambda_{QCD}=0.20~ {\rm GeV}$,
and the coupling constant $\alpha_s(m_b)=0.23$.  By the fitting
ground state data mainly, the fixed value of $V_0$ for various
$J^{PC}$ states is listed in TABLE I.

\begin{figure}
\centering
\includegraphics[width=0.3\textwidth]{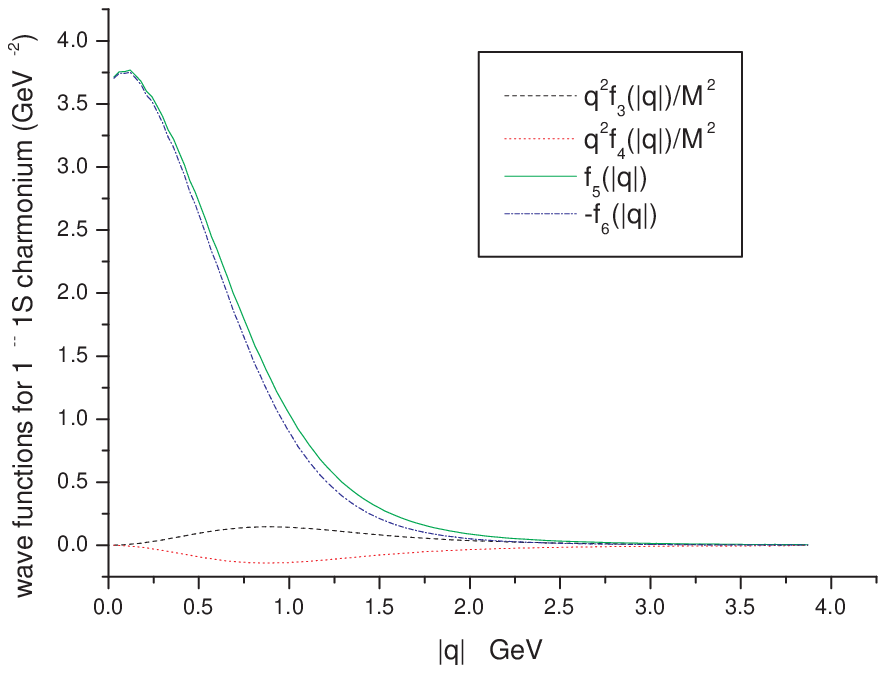}
\includegraphics[width=0.3\textwidth]{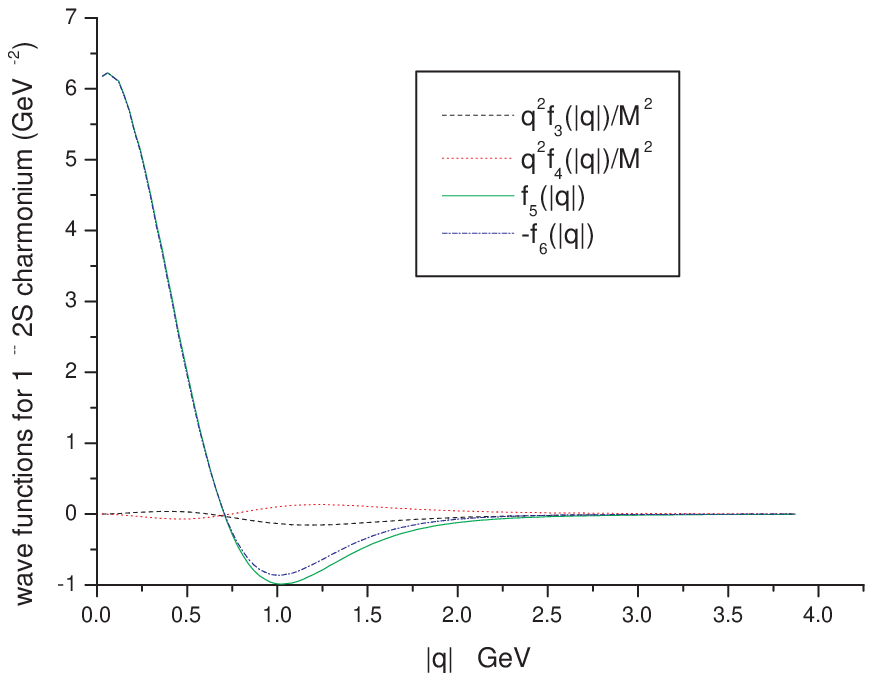}
\includegraphics[width=0.3\textwidth]{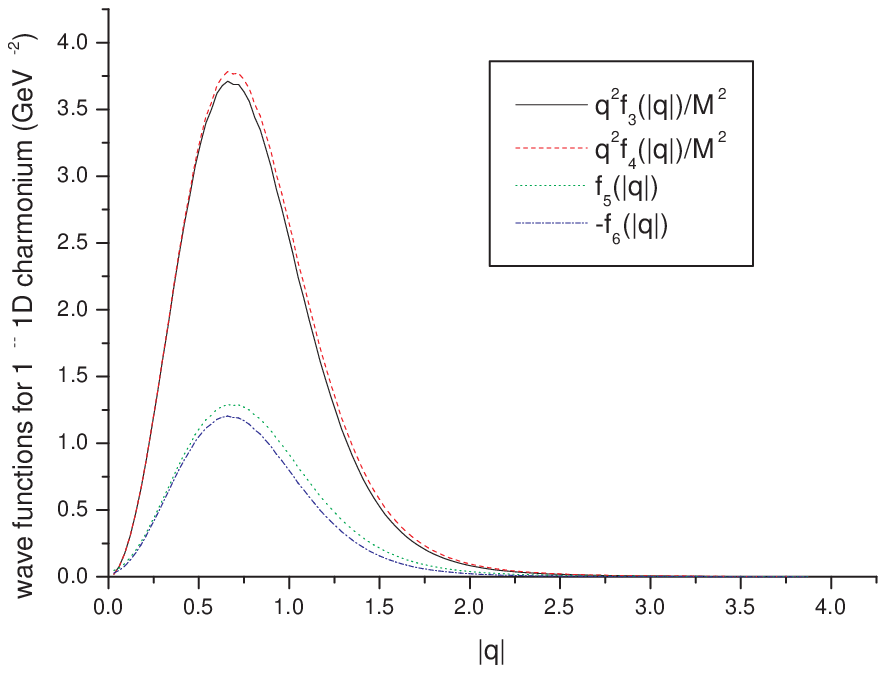}
\includegraphics[width=0.3\textwidth]{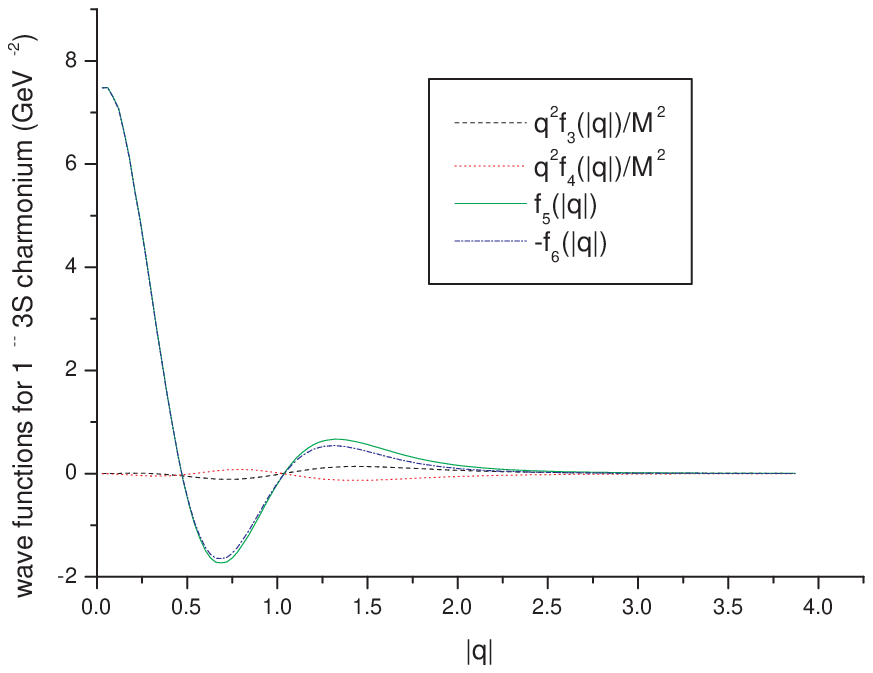}
\includegraphics[width=0.3\textwidth]{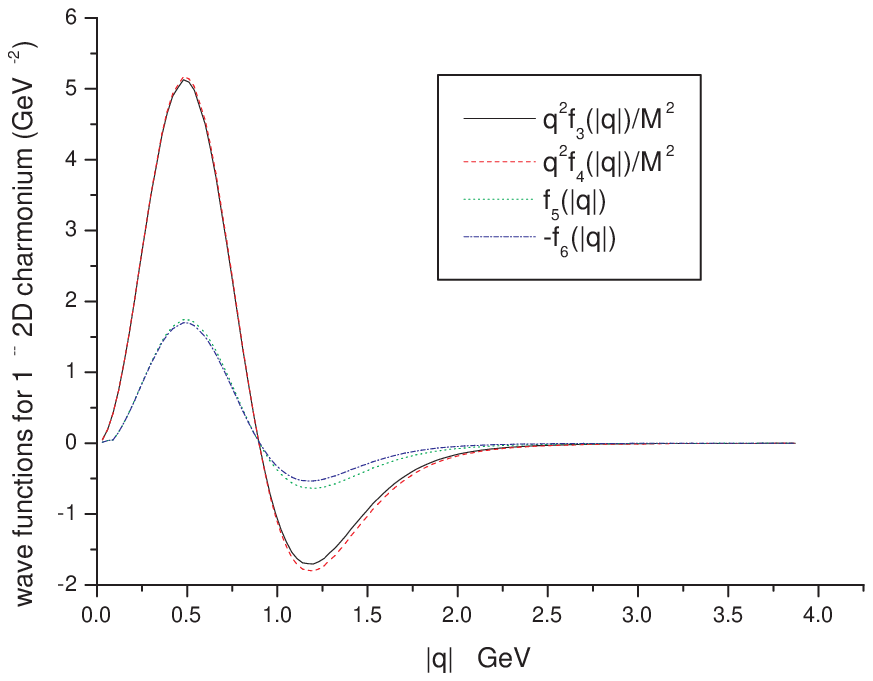}
\caption{The wave functions (solutions) of the low-lying states, the
ground and the first four excited states, (from left to right) for
charmonium with quantum number $J^{PC}=1^{--}$.}
\includegraphics[width=0.3\textwidth]{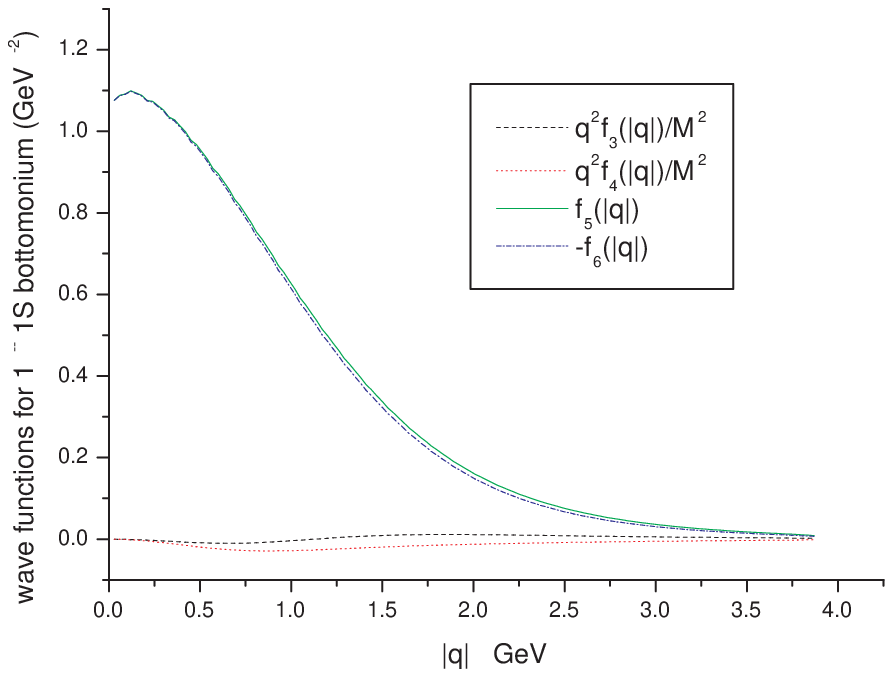}
\includegraphics[width=0.3\textwidth]{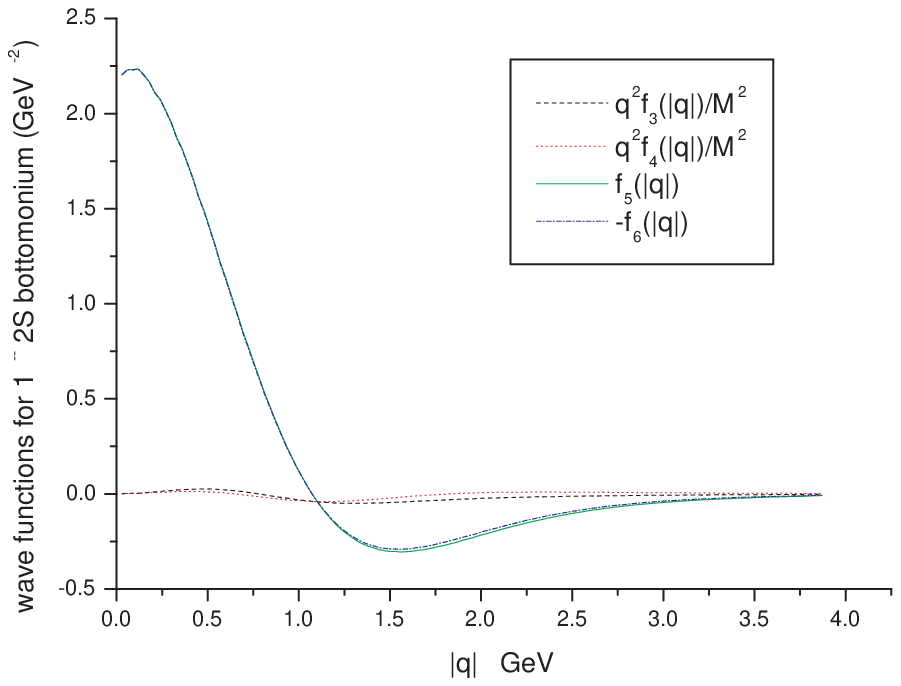}
\includegraphics[width=0.3\textwidth]{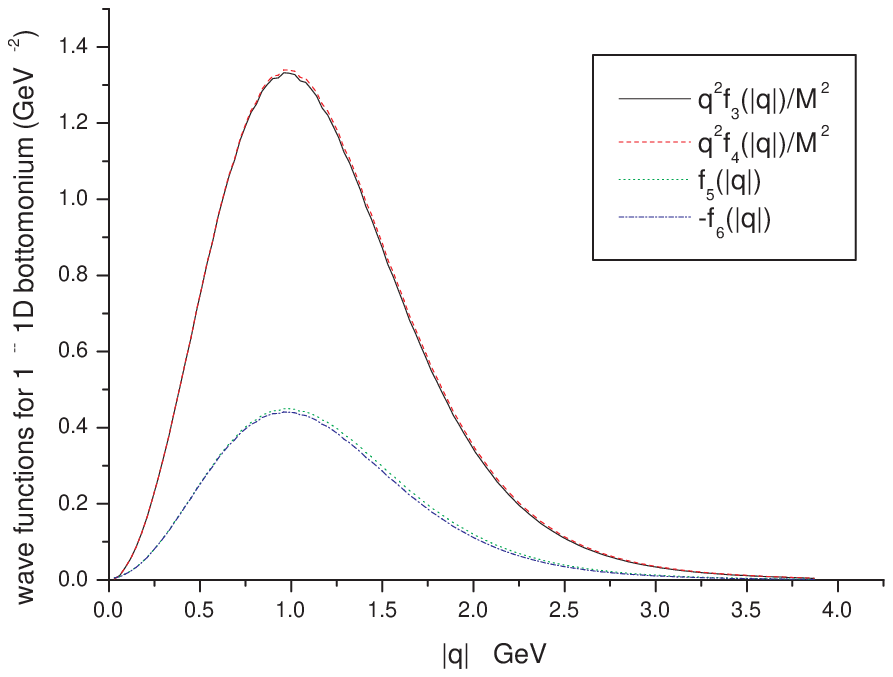}
\includegraphics[width=0.3\textwidth]{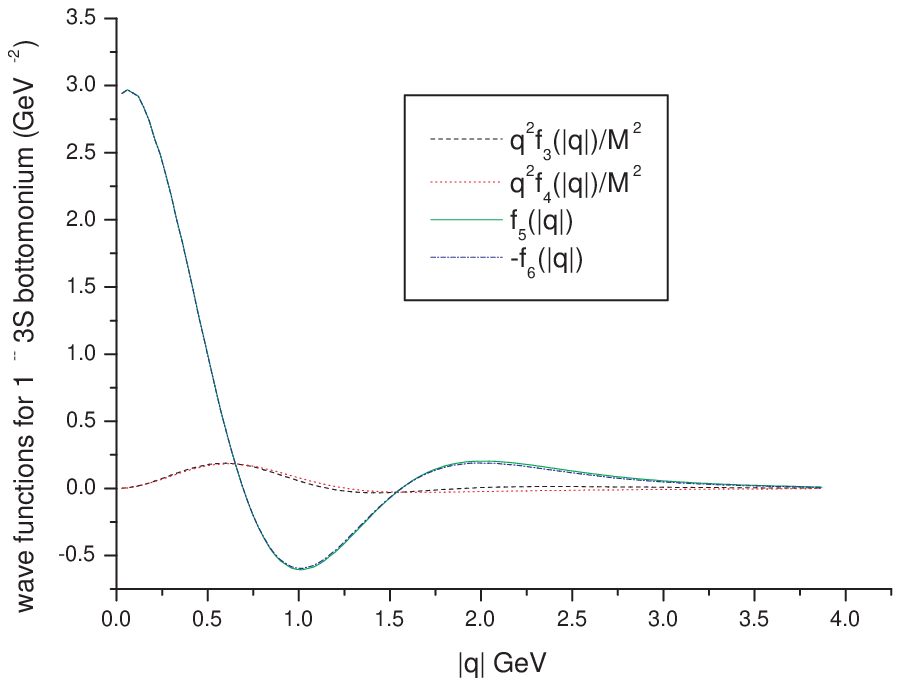}
\includegraphics[width=0.3\textwidth]{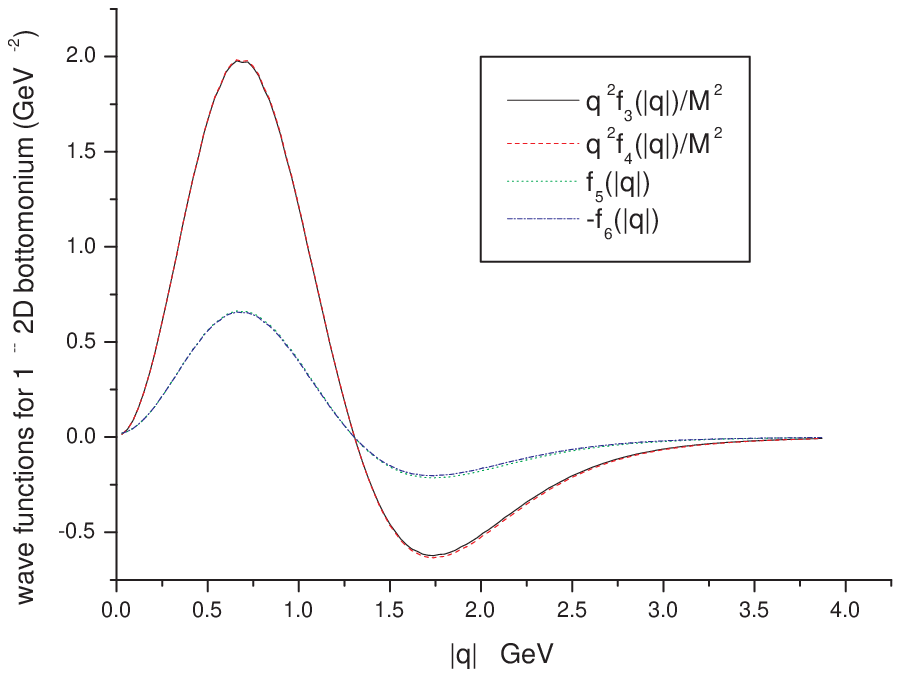}
\caption{The wave functions (solutions) of the low-lying states, the
ground and the first two excited states, (from left to right) for
bottonium with quantum number $J^{PC}=1^{--}$.}
\end{figure}

The spectrum of charmonium and bottomnium (ground states and excited
states) obtained by solving the coupled equations numerically is
shown in TABLE II and TABLE III. Note here that since the
couple-channel effects for the states above the threshold of
`open-charm' or `open-bottom' respectively have not been taken into
account, so the results in TABLE II and TABLE III above the
threshold cannot compare with experimental results directly.

From the tables, one can read out the fine and hyperfine splitting
precisely which are caused by the kernel with the fixed parameters.
Therefore not only the gaps among the excited states and the ground
states with fixed $J^{PC}$ but also the fine and hyperfine splitting
among the states with different $J^{PC}$ are serious tests of the
kernel and the B.S. approach to the heavy quarkonia.

We cannot show all the numerical results of the wave functions which
we have obtained here, alternatively, as typical examples, we only
show some of the obtained wave functions with different quantum
numbers $J^{PC}$ in figures FIGs.1,$\cdots$,8.

As usual cases, from the number of nodes of the wave functions in
the figures we can realize how high an excited one or the ground
each one of the obtained wave functions is.

\begin{figure}
\centering
\includegraphics[width=0.3\textwidth]{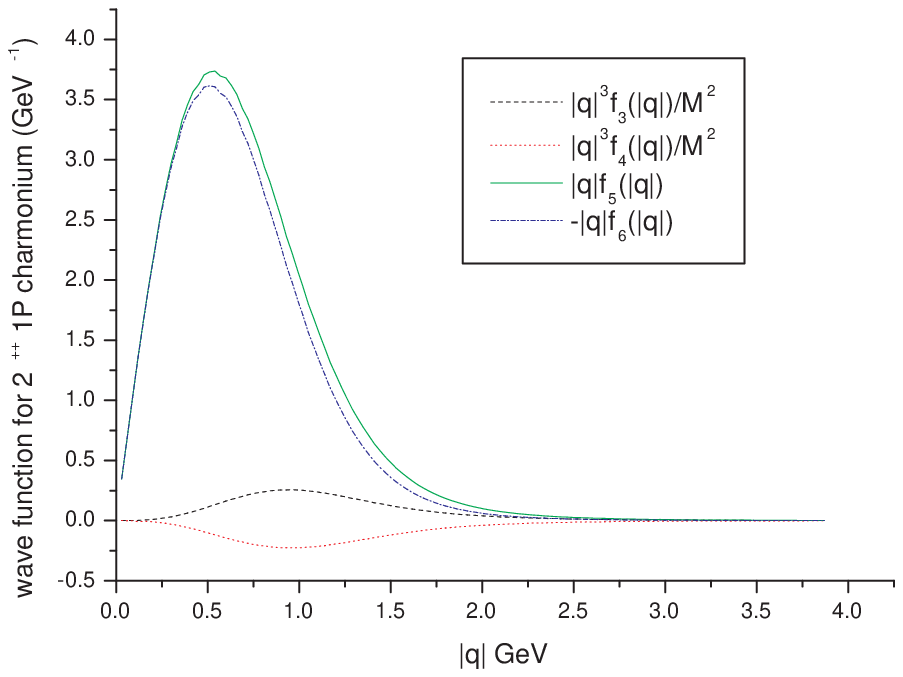}
\includegraphics[width=0.3\textwidth]{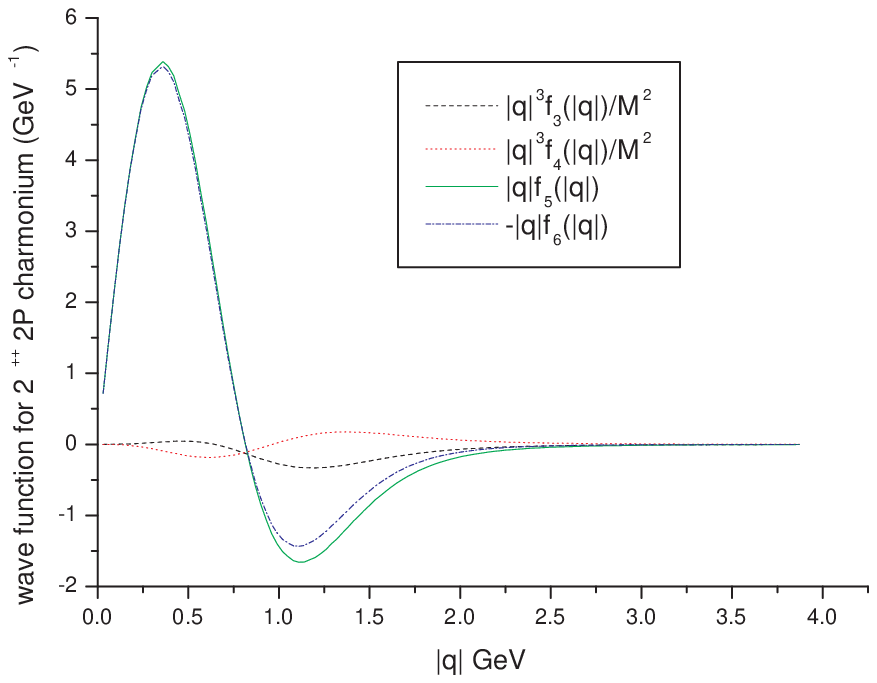}
\includegraphics[width=0.3\textwidth]{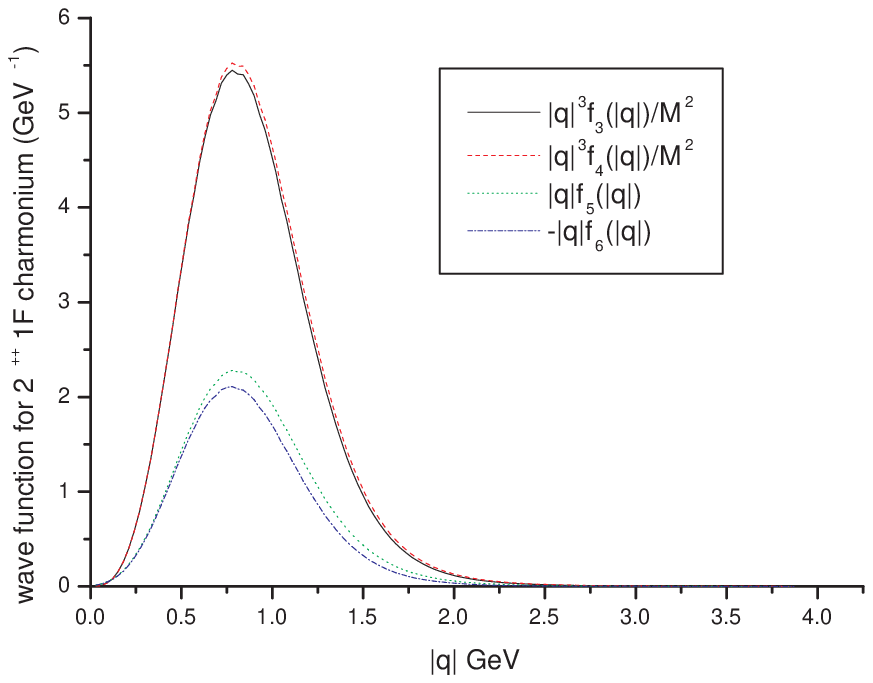}
\includegraphics[width=0.3\textwidth]{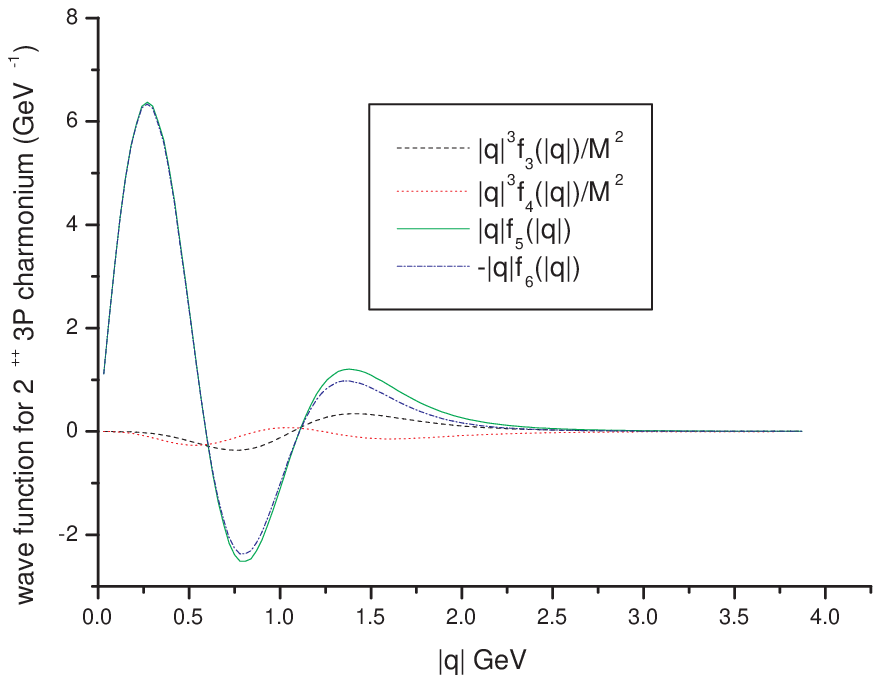}
\includegraphics[width=0.3\textwidth]{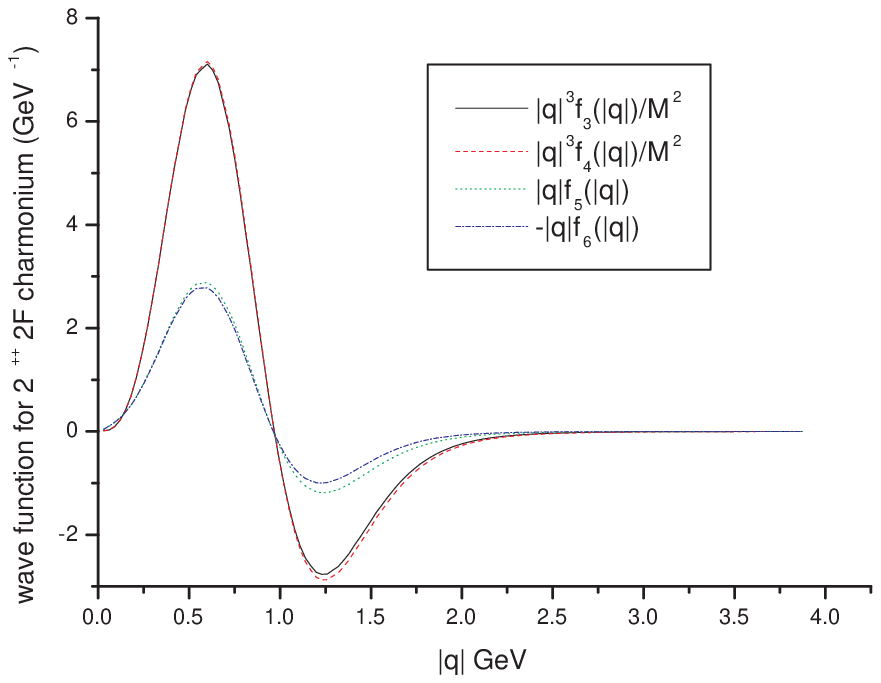}
\caption{The wave functions (solutions) of the five low-lying
states, the ground and the first four excited states, (from left to
right) for charmonium with quantum number $J^{PC}=2^{++}$.}
\includegraphics[width=0.3\textwidth]{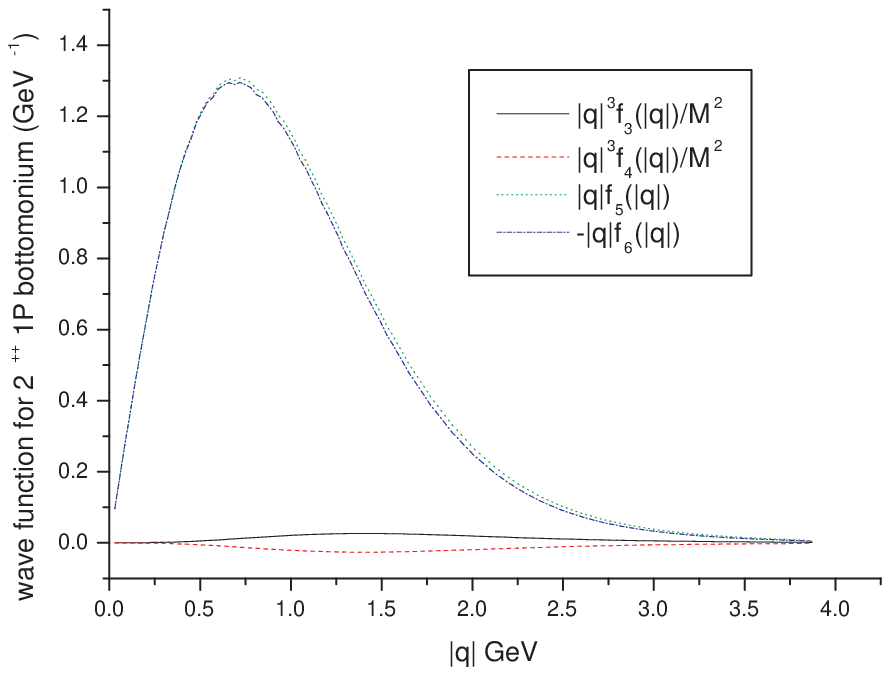}
\includegraphics[width=0.3\textwidth]{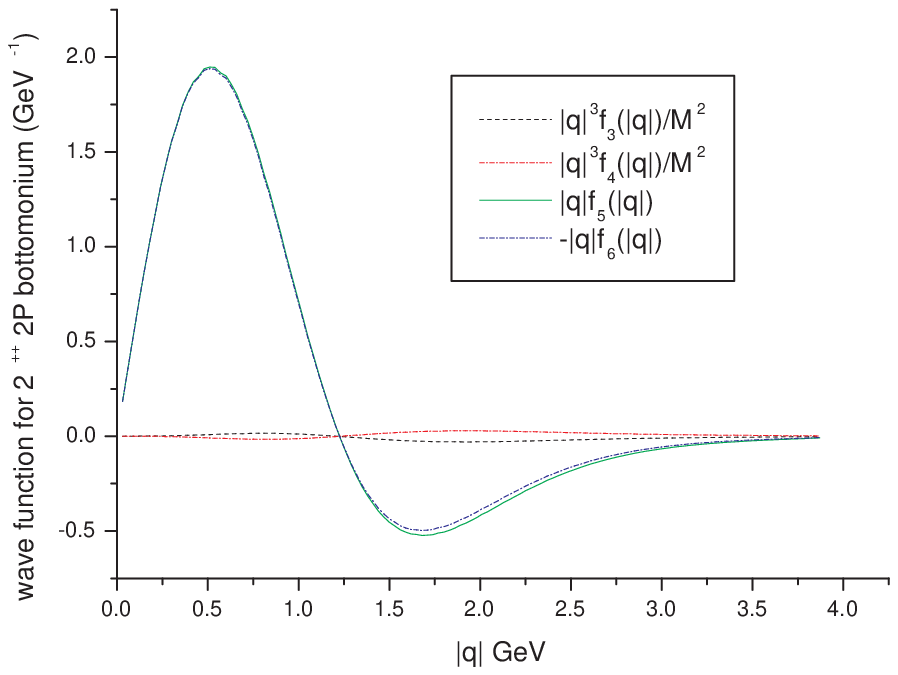}
\includegraphics[width=0.3\textwidth]{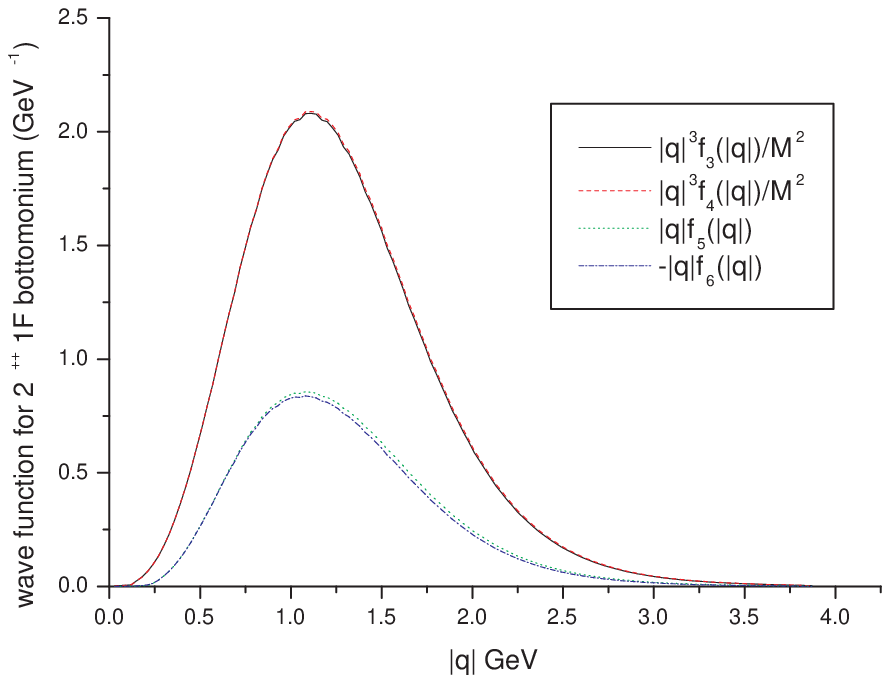}
\includegraphics[width=0.3\textwidth]{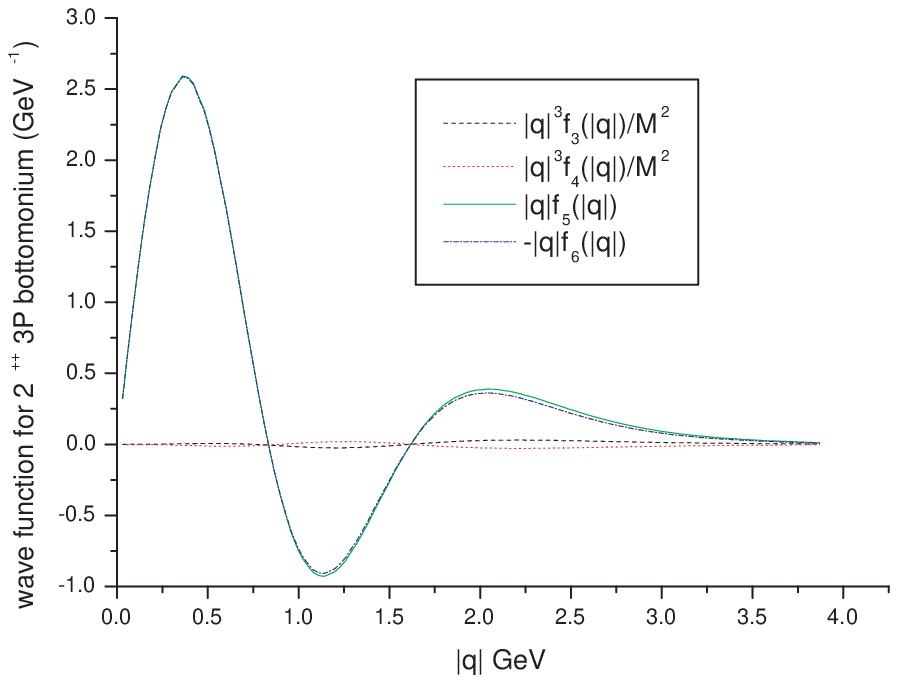}
\includegraphics[width=0.3\textwidth]{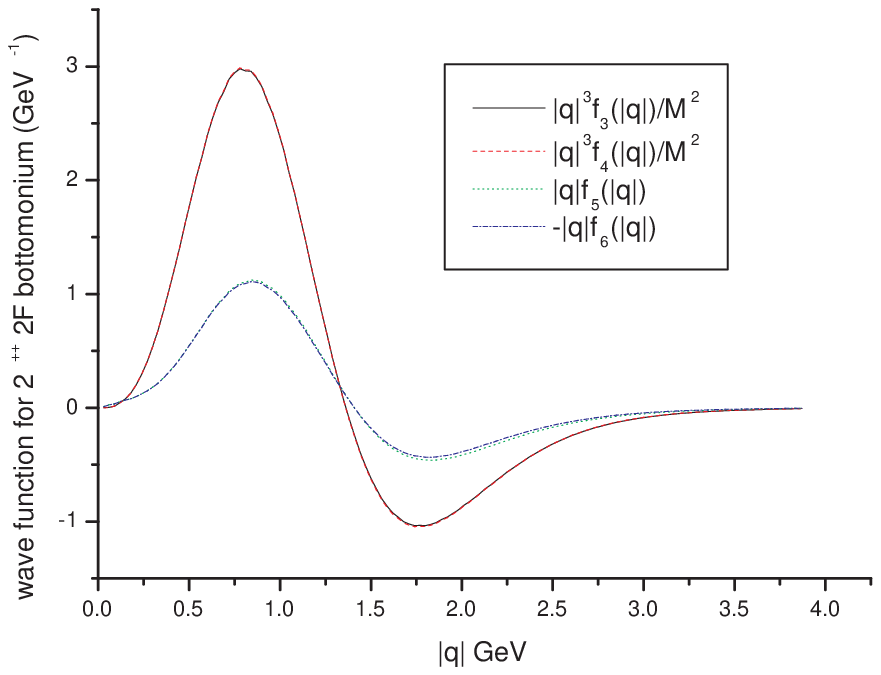}
\caption{The wave functions (solutions) of the five low-lying
states, the ground and the first four excited states, (from left to
right) for bottonium with quantum number $J^{PC}=2^{++}$.}
\label{pdft7}
\end{figure}

From FIG.1, we see clearly that for the states $J^{PC}=0^{-+}$ the
solution has the property $f_1\simeq f_2$, so we may re-write the
wave function Eq.(\ref{EQ:16}) as
\begin{eqnarray}
\varphi_{_{P,0^{-+}}}(q) &=&\left[(1+\frac{\not\!q_{\perp}}{m_1})
{\not\!P}f_1(q) +Mf_2(q)\right]\gamma_{_5}\nonumber\\
&\simeq&\phi_{_{0^{-+}}}(q)\left[(1+\frac{\not\!q_{\perp}}{m_1})
{\not\!P} +M \right]\gamma_{_5}\;,
\end{eqnarray}
here $\phi_{_{0^{-+}}}(q)\simeq f_1(q)\simeq f_2(q)$, the numerical
solution of Eq(\ref{eq0-+}).

From FIGs.2,3,4, we may see that the situations for the states with
quantum numbers $J^{PC}=1^{+-}$, $J^{PC}=0^{++}$ and $J^{PC}=1^{++}$
are similar. Indeed the wave functions of the ground state and the
excited states for $J^{PC}=1^{+-}$ Eq.(\ref{eq1+-0}) become
\begin{eqnarray}
\varphi_{_{P,1^{+-}}}(q)
&=&q_{\perp}\cdot{\epsilon}^{\lambda}_{\perp}\left[
f_1(q)+f_2(q)\Big(1+\frac{\not\!q_{\perp}}{m_1}\Big)\frac{{\not\!P}}{M}
\right]\gamma_5\nonumber\\
&\simeq&
\phi_{_{1^{+-}}}(q)(q_{\perp}\cdot{\epsilon}^{\lambda}_{\perp})\left[1
+\Big(1+\frac{\not\!q_{\perp}}{m_1}\Big)\frac{{\not\!P}}{M}
\right]\gamma_5\;,
\end{eqnarray}
here $\phi_{_{1^{+-}}}(q)\simeq f_1(q)\simeq f_2(q)$, the numerical
solution of Eq.(\ref{eq1+-}); the wave function for $J^{PC}=0^{++}$
Eq.(\ref{eq0++10}) becomes
\begin{eqnarray}
\varphi_{_{P,0^{++}}}(q)&=&
f_1(q){\not\!q_{\perp}}+f_2(q)\frac{{\not\!q_{\perp}}\not\!P}{M}
-\frac{f_1(q)\vec{q}^2}{m_1}\nonumber\\
&\simeq&\phi_{_{0^{++}}}(q)\Big[1+\frac{\not\!P}{M}
+\frac{{\not\!q_{\perp}}}{m_1}\Big]{\not\!q_{\perp}} \,,
\end{eqnarray}
here $\phi_{_{0^{++}}}(q)\simeq f_1(q)\simeq -f_2(q)$, the numerical
solution; the wave functions for $J^{PC}=1^{++}$ Eq.(\ref{eq1++0})
become
\begin{eqnarray}
\varphi_{_{P,1^{++}}}(q) &=&i\varepsilon_{\mu\nu\alpha\beta}
P^{\nu}q_{\perp}^{\alpha}\epsilon^{\beta}\Big[f_1(q)M\gamma^{\mu}+
f_2(q){\not\!P}\gamma^{\mu}
+if_2(q)\varepsilon^{\mu\rho\sigma\delta}
q_{\perp\rho}P_{\sigma}\gamma_{\delta}\gamma_{5}/m_1
\Big]/M^2\nonumber\\
&\simeq& i\phi_{_{1^{++}}}(q)\varepsilon_{\mu\nu\alpha\beta}
P^{\nu}q_{\perp}^{\alpha}\epsilon^{\beta}\Big[M\gamma^{\mu}+
{\not\!P}\gamma^{\mu} +i\varepsilon^{\mu\rho\sigma\delta}
q_{\perp\rho}P_{\sigma}\gamma_{\delta}\gamma_{5}/m_1 \Big]/M^2\,,
\end{eqnarray}
here $\phi_{_{1^{++}}}(q)\simeq f_1(q)\simeq -f_2(q)$, the numerical
solution of Eq.(\ref{eq1++}).

Moreover from FIG.5 and FIG.6 one may see the $S-D$ wave mixing for
the $J^{PC}=1^{--}$ states, and from FIG.7 and FIG.8 one may see the
$P-F$ wave mixing for the $J^{PC}=2^{++}$ states clearly. For the
$J^{PC}=1^{--}$ states, from the figures (FIG.5 and FIG.6) we can
see that for the first two states (the solutions for the ground one
and the first excited one) the $S$-wave components $f_5$ and $f_6$
are dominant, and for the third state (the solution for the second
excited one) the $D$-wave components $f_3$ and $f_4$ are dominant,
etc. Therefore in TABLE I, we denote the first two states as
$J^{PC}(^3S_1)=1^{--}$ and the third one as $J^{PC}(^3D_1)=1^{--}$,
etc in turn for high excited states. For the $J^{PC}=2^{++}$ states,
similarly we can see from the figures that the $P$-wave components
$f_5$ and $f_6$ are dominant in the first two states (the ground one
and the first excited one), and the $F$-wave components $f_3$ and
$f_4$ are dominant in the third state (the second excited state),
etc. Therefore in TABLE II, we denote the first two states as
$J^{PC}(^3P_2)=2^{++}$ and the third one as $J^{PC}(^3F_2)=2^{++}$
etc, in turn for high excited states.

Of the $J^{PC}=1^{--}$ states we may also see from the figure that
for the $S$-wave dominant states they have the properties:
$f_3\simeq -f_4\equiv \phi_{_{1^{--}}}(q)$ and $f_5\simeq -f_6\equiv
\psi_{_{1^{--}}}(q)$, so the solutions (wave functions) can be
re-written (from Eq.(\ref{eq1330})) as
\begin{eqnarray}\label{eq13301}
\displaystyle\varphi_{_{P,1^{--}}}^{\lambda}(q_{\perp})
&\simeq&\phi_{_{1^{--}}}(q)(q_{\perp}\cdot{\epsilon}^{\lambda}_{\perp})\Big[\Big(\frac{-q^2}{Mm_1}+
\frac{{\not\!q}_{\perp}}{M}\Big)-
\frac{{\not\!P} {\not\!q}_{\perp}}{M^2}\Big] \nonumber\\
&+&\displaystyle\psi_{_{1^{--}}}(q)\Bigg\{\Big(
M{\not\!\epsilon}^{\lambda}_{\perp}
+q_{\perp}\cdot{\epsilon}^{\lambda}_{\perp}\frac{M}{m_1}\Big)
-\Big[{\not\!\epsilon}^{\lambda}_{\perp}{\not\!P}
+\frac{{\not\!P}(q_{\perp}\cdot{\epsilon}^{\lambda}_{\perp})}{m_1}
-\frac{({\not\!P}{\not\!\epsilon}^{\lambda}_{\perp}
{\not\!q}_{\perp})}{m_1}\Big]\Bigg\}\;;
\end{eqnarray}
whereas of the $D$-wave dominant states they have $f_3\simeq
f_4\equiv \phi_{_{1^{--}}}(q)$ and $f_5\simeq -f_6\equiv
\psi_{_{1^{--}}}(q)$, so the solutions (wave functions) can be
re-written as
\begin{eqnarray}\label{eqDpd}
\displaystyle\varphi_{_{P,1^{--}}}^{\lambda}(q_{\perp})
&\simeq&\phi_{_{1^{--}}}(q)(q_{\perp}\cdot{\epsilon}^{\lambda}_{\perp})\Big[\Big(\frac{-q^2}{Mm_1}+
\frac{{\not\!q}_{\perp}}{M}\Big)+
\frac{{\not\!P} {\not\!q}_{\perp}}{M^2}\Big] \nonumber\\
&+&\displaystyle\psi_{_{1^{--}}}(q)\Bigg\{\Big(
M{\not\!\epsilon}^{\lambda}_{\perp}
+q_{\perp}\cdot{\epsilon}^{\lambda}_{\perp}\frac{M}{m_1}\Big)
-\Big[{\not\!\epsilon}^{\lambda}_{\perp}{\not\!P}
+\frac{{\not\!P}(q_{\perp}\cdot{\epsilon}^{\lambda}_{\perp})}{m_1}
-\frac{({\not\!P}{\not\!\epsilon}^{\lambda}_{\perp}
{\not\!q}_{\perp})}{m_1}\Big]\Bigg\}\;.
\end{eqnarray}

From the figures (FIG.7 and FIG.8) of the $J^{PC}=2^{++}$ states, we
may see that for the $P$-wave dominant states, the solutions have
the properties: $f_5\simeq -f_6 \equiv \psi_{_{2^{++}}}(q)$ and
$f_3\simeq -f_4 \equiv \phi_{_{2^{++}}}(q)$ and the solutions (wave
functions) can be re-written (from Eq.(\ref{eq2330})) as
\begin{eqnarray}
&\displaystyle\varphi_{_{P,2^{++}}}^{\lambda}(q_{\perp})\simeq
\phi_{_{2^{++}}}(q){\varepsilon}^{\lambda}_{\mu\nu}{q_{\perp}^{\nu}}{q_{\perp}^{\mu}}
\Bigg[\Big(\frac{{\not\!q}_{\perp}}{M}-\frac{q^2}{Mm_1}\Big)-\frac{{\not\!P}
{\not\!q}_{\perp}}{M^2}\Bigg]\nonumber \\
&\displaystyle + \psi_{_{2^{++}}}(q)
{\varepsilon}^{\lambda}_{\mu\nu}{q_{\perp}^{\nu}}\Bigg\{
({\gamma^{\mu}}+\frac{q_{\perp}^{\mu}}{m_1})M-
{\gamma^{\mu}}{\not\!P}-i\frac{1}
{m_1}\epsilon^{\mu\alpha\beta\gamma}
P_{\alpha}q_{\perp\beta}\gamma_{\gamma}\gamma_{5}\Bigg\}\,;
\end{eqnarray}
whereas for the $F$-wave dominant states, the solutions have the
properties: $f_5\simeq f_6\equiv  \psi_{_{2^{++}}}(q)$ and
$f_3\simeq f_4\equiv \phi_{_{2^{++}}}(q)$ and the solutions (wave
functions) can be re-written (from Eq.(\ref{eq2330})) as
\begin{eqnarray}
&\displaystyle\varphi_{_{P,2^{++}}}^{\lambda}(q_{\perp})\simeq
\phi_{_{2^{++}}}(q){\varepsilon}^{\lambda}_{\mu\nu}{q_{\perp}^{\nu}}{q_{\perp}^{\mu}}
\Bigg[\Big(\frac{{\not\!q}_{\perp}}{M}-\frac{q^2}{Mm_1}\Big)+\frac{{\not\!P}
{\not\!q}_{\perp}}{M^2}\Bigg]\nonumber \\
&\displaystyle + \psi_{_{2^{++}}}(q)
{\varepsilon}^{\lambda}_{\mu\nu}{q_{\perp}^{\nu}}\Bigg\{
({\gamma^{\mu}}+\frac{q_{\perp}^{\mu}}{m_1})M-
{\gamma^{\mu}}{\not\!P}-i\frac{1}
{m_1}\epsilon^{\mu\alpha\beta\gamma}
P_{\alpha}q_{\perp\beta}\gamma_{\gamma}\gamma_{5}\Bigg\}\,.
\end{eqnarray}

Finally we would like to discuss the wave mixture further. As shown
in TABLE.II and Eqs.(\ref{eq13301},\ref{eqDpd}), each of the states
(either charmonium or bottomonium) for $J^{PC}=1^{--}$ contains
$S-D$ wave mixing. Some are $S$-wave dominant and the others are
$P$-wave dominant. The third one of charmonium with $m=3778.9$MeV
below the threshold of `open-charm', for instance, clearly is a
$D$-wave dominant state, and in its decay into $l\bar{l}$, $(l=e,
\mu)$ only its $S$-wave components play a role, so the fraction
width of the pure leptonic decay is comparatively small. Indeed it
corresponds to the observed one $\psi"$ (with mass $m=3772.92$MeV)
as pointed in \cite{Eichten2}. Whereas, similarly there is wave
mixture for bottomonium too, for instance, once more the third one
of bottomonium with $m=10129.5$MeV below the threshold of
`open-bottom' is also a $D$-wave dominant state and it decays into
$l\bar{l}, (l=e, \mu)$ only via its $S$-wave components, so the
fractional width of the pure leptonic decay is comparatively small
too\footnote{In Ref.\cite{Eichten2}, the wave mixture is obtained
via additional interaction and that of charmonium is concerned only,
i.e. the mixture for bottomonium is not discussed, that is different
from here. Here the mixture for charmonium and bottomonium is fully
determined by the B.S. kernel well, so we need to discuss the sense
for bottomonium on experimental observations too.}. Furthermore, the
fractional width of the pure leptonic decay for such a state will be
comparatively much smaller (a quarter) than that of charmonium, due
to the two factors that the charge of bottom-quark is smaller than
that of charm-quark, and the comparative weight of the $S$-wave
component to the $D$-wave component in the $D$-wave dominant state,
which is proportional to $v^2$ ($v_{bottomonium}<v_{charmonium}$) as
indicated in Eq.(\ref{eq1330})), is small. Therefore such a state is
very difficult to be observed either in $e^+e^-$ energy scanning
experiments at CLEO and B-factories (due to low production rate) or
in hadron colliders (due to very small branching ratio for the
lepton pair decay and various backgrounds etc). We believe that all
such $D$-wave dominant states for bottomonium must be still missing
in experiments so far, even if our prediction here is true. We
conjecture that such states may be observed at Z-factory such as
Giga-Z etc elsewhere via $e^+e^-\to (b\bar{b})_{\bf 1^{--}(^3 D_1)}+
\gamma $ or $e^+e^-\to (b\bar{b})_{\bf 1^{--}(^3 D_1)}+ \cdots $,
because there the backgrounds can be controlled comparatively easy,
and numerous such bottomonium states enough for experimental
observation can be produced via on-shell $Z$-boson \cite{cww}. For
the $P-F$ wave mixture, since the first state with quantum numbers
$J^{PC}=2^{++}$ is a `high' excited state already so there are only
fewer of the $J^{PC}=2^{++}$ states below the threshold of
`open-charm' or `open-bottom', thus there are fewer example states
which can be used to test the wave mixture, although the tests and
the situation essentially are quite similar to the cases of $1^{--}$
states for $S-D$ wave mixture.

When the bound states does not consist of a pair of quark and
antiquark (not as charmonium and bottomonium here), the quantum
number $C$ is not a good one, then the present way to solve the
problem (the relevant B.S. equation) should be changed accordingly,
but its main steps may be still kept and interesting results, which
are different from the present, are obtained finally. In fact we
have considered the double heavy system ($c\bar b$) or ($\bar{c}b$)
as an example for non-(quark-antiquark) binding system and solved
the relevant B.S. equation in a similar way, but due to differences
we put the results and discussions about the double heavy system
($c\bar b$) or ($\bar{c}b$) elsewhere in Ref.\cite{glzx}.

\vspace{2mm} \noindent {\bf Acknowledgement} This work was supported
in part by the National Natural Science Foundation of China (NSFC)
under Grant No.10847001, No.10875155, No.10675038 and No.10875032.
This research was also supported in part by the Project of Knowledge
Innovation Program (PKIP) of Chinese Academy of Sciences, Grant No.
KJCX2.YW.W10.


\begin{thebibliography}{99}

\bibitem{Eichten1} E. Eichten, K. Gottfried, T. Kinoshita, J.B.
Kogut, K.D. Lane and T.-M. Yan, Phys. Rev. Lett. {\bf 34}, (1975)
369 [Erratum-ibid. {\bf 36}, (1976) 1276];  E. Eichten, K.
Gottfried, T. Kinoshita, K.D. Lane and T.-M. Yan, Phys. Rev. {\bf
D17}, (1978) 3090  [Erratum-ibid. {\bf D21}, (1980) 313].
\bibitem{Eichten2} E. Eichten, K. Gottfried, T. Kinoshita, K.D.
Lane and T.-M. Yan, Phys. Rev. {\bf D21}, (1980) 203.

\bibitem{Godfrey} Stephen Godfrey and Nathan Isgur,
Phys. Rev. {\bf D32},  (1985) 189; Stephen Godfrey and Richard
Kokoski, Phys. Rev. {\bf D43}, (1991) 1679; N. Isgur, D. Scora, B.
Grinstein and M. B. Wise, Phys. Rev. {\bf D39},  (1989) 799

\bibitem{ypk} Yu-Qi Chen and Yu-Ping Kuang, Phys. Rev. {\bf D46},
(1992) 1165, Erratum-ibid {\bf D47}, (1993) 350.

\bibitem{ktchao} Ji-Zhong Lou, Dan-Hua Qin, Yi-Bing Ding and Kuang-Ta
Chao, Commun. Theor. Phys. {\bf 30}, (1998) 93; Cong-Feng Qiao,
Han-Wen Huang and Huang-Ta Chao, Phys. Rev. {\bf D54}, (1996) 2273.

\bibitem{caix} C.B. Yang and X. Cai, Phys. Rev. {\bf D51}, (1995)
6332.

\bibitem{rob}
J. Zeng, J.W. Van Orden and C. D. Roberts, Phys. Rev. {\bf D52},
(1995) 5229.

\bibitem{BS}
E. E. Salpeter and H. A. Bethe, Phys. Rev. {\bf 84}, (1951) 1232.

\bibitem{salp}
E. E. Salpeter, Phys. Rev. {\bf 87}, (1952) 328.

\bibitem{chenjk}
Chao-Hsi Chang and Jao-Kai Chen, Commun. Theor. Phys. {\bf 44} (2005)
646-650.

\bibitem{cskimwang}C. S. Kim and Guo-Li Wang, Phys. Lett. {\bf B584}
(2004) 285.

\bibitem{changwang}Chao-Hsi Chang, Jiao-Kai Chen, Xue-Qian Li and Guo-Li Wang,
 Commun. Theor. Phys. {\bf 43} (2005) 113.

\bibitem{rosner}
J. Rosner, Comm. Nucl. Part. Phys. {\bf 16}, 109 (1986).

\bibitem{wise}
N. Isgur, M. B. Wise, Phys. Rev. {\bf D43}: 819 (1991).

\bibitem{wang1}Guo-Li Wang,
Phys. Lett. {\bf B633} (2006) 492.

\bibitem{wang2}Guo-Li Wang,
Phys. Lett. {\bf B650} (2007) 15.

\bibitem{PDG}Particle Data Group, Phys. Lett. {\bf B667} (2008) 1.

\bibitem{etab}BABAR Collaboration, Phys. Rev. Lett. {\bf 101} (2008) 071801.


\bibitem{resag}J. Resag and C.R. M\"{u}nz, Nucl. Phys. {\bf A590} (1995)
735.

\bibitem{cww}Chao-Hsi Chang, Jian-Xong Wang and Xing-Gang Wu, in
preparation.

\bibitem{glzx}Chao-Hsi Chang and Guo-Li Wang, in
preparation.

\end{thebibliography}
\end{document}